# Second-Order Defeaturing Estimator of Manufacturing-Induced Porosity on Structural Elasticity

Shiguang Deng[a, *], Carl Soderhjelm[a], Diran Apelian[a], Krishnan Suresh[b]

[a] ACRC, Materials Science and Engineering, University of California, Irvine, CA, USA

[b] Mechanical Engineering, University of Wisconsin, Madison, WI, USA

**Abstract**

Manufactured metallic components often contain non-uniformly distributed pores of complex morphologies. Since such porosity defects have a significant influence on material behaviors and affect the usage in high-performance applications, it is significant to understand the impact of porosity characteristics on the behaviors of components. In this work, a gradient-enhanced porosity defeaturing estimator, which allows for the modeling of pore geometry and spatial distribution, is proposed within a general elastostatic framework. In this approach, the first-order shape sensitivity is implemented to account for the change in the elastic quantity of interests to variations of pore sizes and shapes, which is then supplemented by a second-order shape sensitivity whose mixed partial derivative quantifies the interactions between pores in proximity. The efficacy of the proposed method comes from its posterior manner that it only relies on field solutions of reference models where pores are suppressed. In this context, meshing difficulty and solution convergence issue are avoided, which would otherwise arise in a direct finite element analysis on porous structures. The impact of porosity on structural elastic performance is approximated using a second-order Taylor expansion where the topological difference between the porous and reference domains is estimated by topological sensitivity; the field variables on pore boundaries are approximated as explicit functions of design variables using exterior formulations. Numerical results show that the elastic performances of components are influenced by the existence of pores. The pore-to-pore interactions are significant when pores are close by.

## 1. INTRODUCTION

Porosity is a common process-induced defect in many metallic components manufactured using different technologies such as high pressure die casting (HPDC) [1] and direct laser melting (DLM) [2]. In HPDC process, porosity can be formed both as macro-porosity where air is rejected from molten metal and entrapped in the mold as spherical gas bubbles, and/or as micro-shrinkage which occurs in regions with inadequate liquid metal flow [3–5]. On the other hand, the generation of irregularly shaped pores in DLM is a process-induced phenomenon that strongly depends on various parameters such as inadequate fusion, incomplete remelting, low powder packing density, and mismatch of powder morphologies between different layers [6–9]. Manufacturing techniques lead to different pore defects in the form of morphology (e.g., pore size, shape, and topology) and spatial distribution (e.g., the nearest distance between neighbors). The porosity defect has a major

---

*Corresponding author.

Email address: sdeng9@wisc.edu (Shiguang Deng)

influence on the resultant mechanical properties. The need to incorporate the role of porosity in the analysis of manufactured components is pivotal.

Synthetic models of primitive geometries are often utilized to represent pore morphologies within matrix materials for simplicity. Waters [10] used spherical voids to represent pore geometries in magnesium alloys to characterize the void behaviors during damage accumulation. Prithivirajan and Sangid [11] employed spherical voids to investigate the critical pore size during dynamic loading conditions (metal fatigue) on selective laser melting (SLM) nickel-based alloys. Recent advances in multiscale methods coupled with synthetic pore models allow to correlate local material porosity defects with global structural performance. Ghosh et al. [12] integrated ellipsoid-shaped pores into an adaptive multi-level model with varying resolutions on each scale. While this model included three levels (macro, meso, and microscopic analysis), coupling between different levels was accomplished through asymptotic homogenization. Considering that pores in nickel-based cast alloys have complex morphologies with convex and concave radii, Taxer et al. [13] developed a more sophisticated synthetic model by intersecting three identical perpendicular ellipsoids at a common geometric center. This model could represent pore spatial ramifications by adjusting aspect ratios of ellipsoids. Deng et al. developed a concurrent multiscale model in [14] to predict the impacts of spatially varying porosity on engineered alloys. Their model integrates with a microstructure reconstruction algorithm to allow for complex microscale morphologies with various pore descriptors including pore volume fraction, number, size, geometry, and neighbor distances. However, multilevel models generally require significant memory allocation and high computational costs due to frequent data exchange and coupled equilibrium equations between scales [15].

Actual porosity characteristics can be captured by computed tomography (CT), a non-destructive detection method for the analysis of material internal defects. Pore morphology and distribution are reconstructed in 3D models through image processing of scanning data [15,16]. With CT as an enabling technology, researchers have tried to incorporate reconstructed porosity characteristics with finite element (FE) models, followed by a direct solve. For example, actual shapes and sizes of casting pores were characterized by light microscopy and reproduced in a FE model to correlate with local stress concentrators in [18]. Pore geometries reconstructed from tomography scans were integrated with a micromechanical model to predict their influence on elastoplastic behaviors [19]. A quantitative description of non-uniformly distributed pores was incorporated into a 3D micromechanical analysis in [20]. It was found that the distributions of local stress depend on pores' size, orientation, and spatial arrangement. Hangai and Kitahara [21] distinguished gas and shrinkage-induced pores by fractal analyses in terms of individual shapes and congregated spatial distributions. It seems straightforward to solve porosity problems using actual geometries; however, numerical issues can arise; the presence of small-sized and irregularly shaped pores significantly increases meshing complexity and computational expense in FE models. As an example, let's consider the W-profile plate produced by HPDC which contains more than 1300 pores represented by red dots in Figure 1(a). In this example, small elements are generated in the vicinity of pore surfaces to address geometric adaptivity and element size transition, which not only results in a substantial mesh size but also raises element quality issues. For example, distorted or inverted elements [22] can be observed close to pore boundaries as shown in Figure 1(c). Ill-shaped elements can further deteriorate the global stiffness matrix in FE systems, leading to slow convergence or even failure [23].

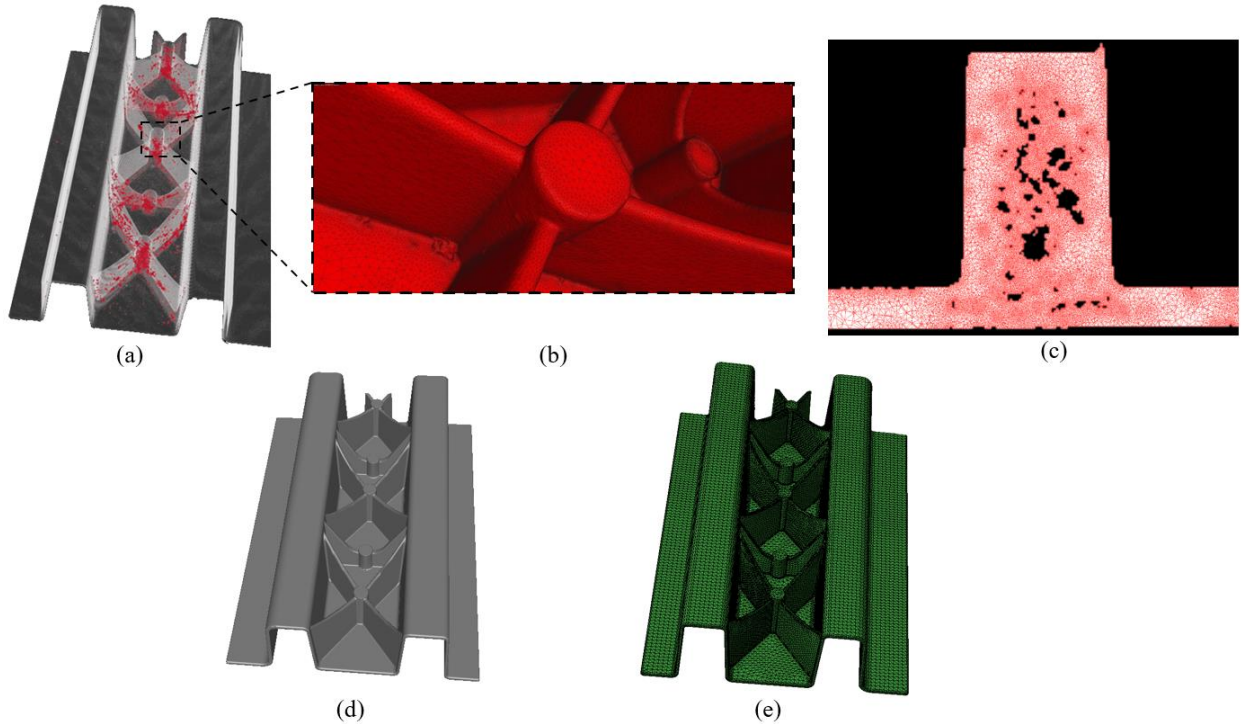

**Figure 1:** (a) Manufactured metallic component with more than 1300 pores whose morphologies and spatial distributions are reconstructed from 3D X-ray tomography; (b) an extremely fine FE mesh is created with more than 6.2 million tetrahedrons is created to capture the porous part's geometric details; (c) a cross-sectional view of the FE mesh illustrates the existence of irregularly shaped pores in proximity and ill-shaped elements near pore surface; (d) a simplified (i.e., defeatured) model without any pores; and (e) a sufficiently fine mesh on the defeatured model only requires less than 0.25 million elements, i.e., about 4% of its porous counterpart.

Recent development in defeaturing techniques provide ways to efficiently integrate dimensional characteristics of pores. Defeaturing is a model simplification approach to preserve certain model properties during geometrical changes. Its basic idea is to work on a simplified model after the removal of features and the performance difference (defeaturing error), between the fully-featured and defeatured models, is estimated via a posterior estimator. During defeaturing, various design sensitivities are utilized. Specifically, shape sensitivity can be used to compute the change in an arbitrary quantity of interest when models' boundary is subject to infinitesimal perturbations [24]; whereas topological sensitivity approximates the functional changes when an infinitesimally small hole is introduced onto a domain [25]. Li and Gao [26] estimated the error of suppressing arbitrarily sized geometric features by using adjoint theory where the feature can be negative (i.e., geometry cutout) or positive (i.e., geometry addition). In their later work [27], a second-order defeaturing method was proposed for error estimations when multiple interactive boundary features are suppressed. A limitation of this method is that it only accounts for boundary features but not for topological changes. Gopalakrishnan and Suresh [28] quantitatively analyzed the interval of defeaturing errors, whose bounds were approximated through monotonicity analysis. Turevsky et al. [29] developed a feature sensitivity by integrating shape sensitivities over geometric transformations to estimate the performance functions in a 2D Poisson problem where topological change was approximated by the shape sensitivity. In a recent work by Deng et al. [30], a first-order defeaturing estimator was developed to exploit the morphology of manufacturing-induced pores. To estimate the influence of pores on part behaviors, topological and shape sensitivities were systematically integrated where each pore was assumed far away from neighbors so that interactions were neglected.

Despite many efforts to incorporate the influence of porosity characteristics on structure performance, shortcomings and gaps remain.

1) Previous defeaturing methods mainly focus on removing one single isolated feature (e.g., void) from the design domain [27–29]. Since each feature is assumed as one term in a linear Taylor expansion, the effect of removing multiple features is addressed by simple additions without considering interactions among them. However, on manufactured components, pores are often clustered nearby, as shown in Figure 1(a) and (c). This dramatically increases pore-to-pore interactions and invalidates the presumed '*isolated*' status. A more sophisticated approach is necessary to account for these interactions.

2) Synthetic pore models are not capable of representing actual pore morphologies, and multiscale methods are computationally expensive due to hierarchical material modeling on different scales. On the other hand, direct FE approaches that explicitly exploit tomography reconstructed porosity characteristics often experience difficulties in addressing meshing and convergent issues. There is a need for a computationally *efficient* framework to correlate tomography-enabled 3D quantitative porosity data with structural performance.

3) Design sensitivities, e.g., size, shape, or topological sensitivity, are mainly integrated into frameworks of structural optimization by providing gradient information [23,24,30]. To our best knowledge, this work is the first to utilize higher-order design sensitivities to perform quantitative analysis on pore interactions.

Inspired by previous defeaturing methods, a novel second-order porosity-oriented estimator is proposed to predict metallic part behaviors due to porosity intrinsic characteristics. Its numerical advantage can be explained by the W-profile plate example in Figure 1(d) where the proposed method only requires solutions on a pore-free structure whose mesh size and quality are much improved as illustrated in Figure 1(e). Moreover, design sensitivities will be derived in boundary forms that only require the 2D surface mesh of pores, dramatically reducing computational costs compared to a classic FE analysis on 3D volume elements.

The remainder of the paper is organized as follows. In Section 2, we pose the problem statement and provide an overview of the proposed approach. We review background knowledge in Section 3. Our proposed method is detailed in Sections 4-6, and it is examined by numerical experiments in Section 0. Conclusion and future works are provided in Section 8.

## 2. PROBLEM STATEMENT

In this section, we formulate the porosity problem through mathematical statements, and provide a general overview of our proposed method, which, different from single feature suppression methods in [28,31], provides a second-order defeaturing methodology to account for the impacts of pore-to-pore interactions on structural elastic performances. Before proceeding to technical details, a summary of critical mathematical symbols and their meanings are given in Table 1. Based on this table, the particular meaning of symbols should be understood from the context.

Table 1: Critical mathematical symbols and their meanings.

| Symbol | Meaning | Symbol | Meaning |
|---|---|---|---|
| $\Omega$ | Geometry domain | **K** | FE stiffness |
| $\Gamma$ | Domain boundary | **f** | External load |
| $\Psi$ | Performance function | $\mathcal{L}$ | Lagrangian equation |
| $\eta, t, s$ | Shape parameters | **P, Q** | Lagrangian multipliers |
| $\xi$ | Topological parameter | $\mathcal{T}$ | Sensitivity field |
| **n** | Normal direction | $\mathcal{D}$ | Estimator value |
| **x** | Points in current configuration | **T** | Shape transformation |
| **X** | Points in reference configuration | **F** | Transformation tensor |
| **z** | Primary solution | **V, W** | Design speeds |
| $\lambda$ | Adjoint solution | **J** | Jacobian matrix |
| g | Generic function | $I_D$ | Effectivity index |
| $\sigma$ | Stress | U | Strain energy |
| $\varepsilon$ | Strain | E | External work |
| $\mathbb{C}$ | Tangent moduli | $\Pi$ | Total potential energy |

## 2.1 Mathematical formulation

Consider an arbitrary 3D domain with two irregularly shaped pores in the close distance as in Figure 2(a). This porous domain, considered as the difference between the dense (or reference) domain ($\Omega$) and the two pores ($\Omega_{pi}$ and $\Omega_{pj}$), can be expressed as $\Omega$-$\Omega_{pi}$-$\Omega_{pj}$ in Figure 2(b). For the sake of simplification, we assume sizes of pores ($\|\Omega_p\|$) are much smaller than that of domain ($\|\Omega\|$)), pores are far from domain surfaces ($\Gamma$) to prevent surface effects, and the inter-pore distance is so small that pore-to-pore interactions cannot be neglected, as stated in Equation (2.1).

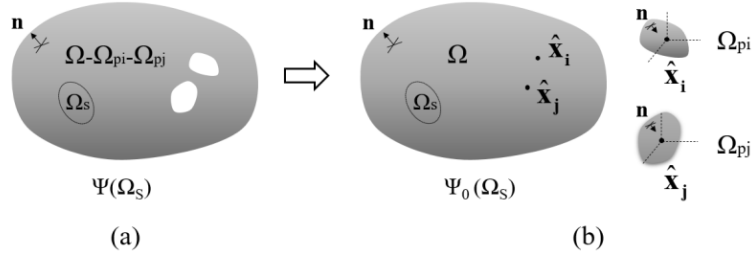

Figure 2: Schematic representation of an arbitrary domain with irregular pores. (a) The porous domain with two closely distanced pores, (b) The porous domain is equivalently viewed as the difference between a reference (dense) domain and the two pores.

$$\begin{cases} \|\Omega_P\| \ll \|\Omega\| \\ \|\Omega_P\| \ll \text{distance}(\Gamma, \Gamma^p) \\ \text{distance}(\Gamma^{pi}, \Gamma^{pj}) \ll \text{distance}(\Gamma, \Gamma^p) \end{cases} \quad (2.1)$$

To compute the displacement solutions (**z**) on a porous domain, a boundary value problem (BVP) needs to be solved:

$$\begin{cases} -\nabla \cdot \boldsymbol{\sigma}(\mathbf{z}) = \mathbf{f}^{\mathbf{b}} & \mathbf{x} \in \Omega - \Omega_{pi} - \Omega_{pj} \\ \mathbf{z} = \hat{\mathbf{z}} & \mathbf{x} \in \Gamma^{h} \\ \boldsymbol{\sigma}(\mathbf{z}) \cdot \mathbf{n} = \mathbf{f}^{\mathbf{s}} & \mathbf{x} \in \Gamma^{s} \\ \boldsymbol{\sigma}(\mathbf{z}) \cdot \mathbf{n} = 0 & \mathbf{x} \in \Gamma^{pi}, \Gamma^{pj} \end{cases} \quad (2.2)$$

where material points ($\mathbf{x}$) on the porous domain are subject to the body force $\mathbf{f}^{\mathbf{b}}$, a displacement field $\hat{\mathbf{z}}$ is prescribed over the Dirichlet boundary $\Gamma^{h}$, a surface traction $\mathbf{f}^{\mathbf{s}}$ is applied on the Neumann boundary $\Gamma^{s}$ with a unit outward normal vector $\mathbf{n}$, zero Neumann conditions are applied on the pore surfaces ($\Gamma^{pi}$ and $\Gamma^{pj}$) associated with the pore $i$ and $j$, and $\boldsymbol{\sigma}$ is the stress field on material points by assuming linear elastic constitutive equation in Equation (2.3).

$$\boldsymbol{\sigma} = \mathbb{C} : \boldsymbol{\varepsilon} \quad (2.3)$$

$$\boldsymbol{\varepsilon} = \frac{1}{2}\left(\nabla \mathbf{z} + \nabla \mathbf{z}^{\mathbf{T}}\right) \quad (2.4)$$

where $\mathbb{C}$ is the linear elastic tangent matrix, and small strain ($\boldsymbol{\varepsilon}$) is computed from the displacement gradients ($\nabla \mathbf{z}$). In addition, we are often interested in a generic performance function defined in the desired region $\Omega_s$ as:

$$\Psi = \iiint_{\Omega_S} g(\mathbf{z}) d\Omega \qquad \Omega_S \subset \Omega - \Omega_{pi} - \Omega_{pj} \quad (2.5)$$

where $g$ represents an arbitrary scalar function dependent on the displacement variable $\mathbf{z}$.

To solve $\Psi(\Omega_s)$ on a porous structure, directly solving Equation (2.2), e.g., through FEM, seems straightforward. But several issues might arise. First, a real casting often contains many pores of complex morphologies. Generating FE volume mesh on such pore surfaces often produces prohibitive meshing costs due to a large number of highly distorted elements. Ill-shaped elements further deteriorate the global stiffness matrix in the iterative solution process, leading to convergence problems. This direct FE approach should be therefore avoided.

On the other hand, if without any pore, it would be much easier to solve FE on the reference domain. Therefore, an alternative method is to develop a posterior estimator to predict the change of performance function in Equation (2.5) when pores are suppressed, i.e., a defeaturing process which considers pores as features in a FE model. Thus, a much simpler BVP can be imposed on the reference domain as:

$$\begin{cases} -\nabla \cdot \boldsymbol{\sigma}(\mathbf{z}) = \mathbf{f}^{\mathbf{b}} & \mathbf{x} \in \Omega \\ \mathbf{z} = \hat{\mathbf{z}} & \mathbf{x} \in \Gamma^{h} \\ \boldsymbol{\sigma}(\mathbf{z}) \cdot \mathbf{n} = \mathbf{f}^{\mathbf{s}} & \mathbf{x} \in \Gamma^{s} \end{cases} \quad (2.6)$$

where the quantity of interest is defined over the same region $\Omega_s$ as:

$$\Psi_0 = \iiint_{\Omega_S} g(\mathbf{z}) d\Omega \qquad \Omega_S \subset \Omega \quad (2.7)$$

From different BVPs in Equation (2.2) and (2.6), two different displacement solutions can be computed, which then results in two different values of performance functions in (2.5) and (2.7). If the quantity of interest is measured in an energy form, the difference (defeaturing error) between the two performance functions is called a global error, otherwise, a goal-oriented error [26]. Estimation of goal-oriented error is more important than its global counterpart since it gives more insights into structural local effects, such as pointwise displacement or local stress concentration. However, goal-oriented error estimation is comparatively more difficult and expensive [26].

2.2 Approach overview

The objective of the proposed method is to estimate the quantity of interests $\Psi(\Omega_s)$ on a porous domain by using its counterpart $\Psi_0(\Omega_s)$ on a reference domain. To build connections between the two performance functions, a domain transformation can be proposed between the porous domain ($\Omega$-$\Omega_{pi}$-$\Omega_{pj}$) and the reference domain $\Omega$ as illustrated in Figure 3.

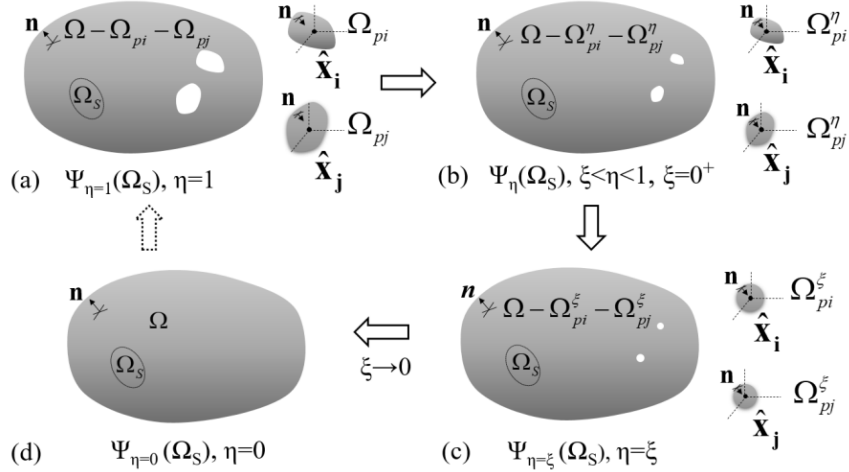

**Figure 3**: Schematic representation of the proposed method. The porosity domains in (a) are sequentially subject to shape perturbations in (b) and topological changes in (c) and are transformed to the reference domain in (d). Our proposed method aims to use the field solutions on the reference domain to approximate any quantity of interests on the original porous domain.

In Figure 3, let the two pores both be parameterized by a continuous geometry parameter $\eta$ on the interval [0, 1]. While the state ($\eta=1$) presents the full-sized pores as in Figure 3(a), the range (0<$\eta$<1) represents a continuously size-shrinking process for the two pores as in Figure 3(b). In this process, the first-order shape sensitivity is employed to account for the variation of performance functions to the shape change of each pore. The sensitivity field is supplemented by a second-order shape sensitivity to quantify pore-to-pore interactions. At ($\eta=\xi$ where $\xi\to0^+$), we assume the two pores are scaled to infinitesimally small sizes ($\xi$) such that the influence of their original shapes can be assumed trivial as in Figure 3(c). If their shapes can be assumed as spheres of equivalent sizes $\xi$, removal of the two small spheres results in the reference domain as in Figure 3(d) where topological sensitivity is utilized to capture the impact of domain topological change. A detailed discussion about the relation between the topological and shape sensitivities can be found in [33].

In summary, the variation (defeaturing error) of the two performance functions in Equation (2.5) of Figure 3(a) and Equation (2.7) of Figure 3(d) can be approximated by exploiting shape and topological sensitivity fields as:

$$\Psi(\Omega_S) - \Psi_0(\Omega_S) = \mathcal{D}_{topo} + \mathcal{D}_{shape} \equiv \mathcal{D}_{pore} \tag{2.8}$$

where $\mathcal{D}_{topo}$ and $\mathcal{D}_{shape}$ are the posterior error estimations from topological and shape sensitivities, respectively. Their combined estimations are coined as the porosity estimator ($\mathcal{D}_{pore}$) in this work.

We have demonstrated in our early work [30] that a first-order estimator may provide sufficiently accurate predictions for sparsely distributed pores where the distance between any two pores (e.g., pore $i$ and $j$) are assumed much larger than their sizes in Equation (2.9). In such a scenario, the porosity estimator $\mathcal{D}_{pore}$ is expressed as the sum of the first-order topological sensitivity ($\mathcal{D}_{topo}$) and shape sensitivity ($\mathcal{D}^1_{shape}$) in Equation (2.10).

$$\text{distance}(\Gamma^{Pi}, \Gamma^{Pj}) \gg \|\Omega_P\| \tag{2.9}$$

$$\mathcal{D}_{pore} = \mathcal{D}_{topo} + \mathcal{D}^1_{shape} \tag{2.10}$$

However, as we have observed in Figure 1 that process-induced pores often agglomerate in certain regions, the close distances between pores violate the presumption in Equation (2.9) and cast doubts on the accuracy of the linear estimator. Our study, therefore, aims to answer the following critical questions:

1) *Does the short inter-pore distance significantly affect the estimation accuracy of the linear porosity estimator in [30]? If yes, how to mathematically quantify such interactions in the proposed second-order estimator?*

2) *How much accuracy improvement does the new estimator achieves compared to its linear counterpart? And what parameters would affect its accuracy?*

3) *Can we use the second-order shape sensitivity strategy in [27] to solve our porosity problem? And what are the major differences between [27] and our method?*

Answers to the above questions are theoretically investigated in method derivations and numerically demonstrated by experiments. In the next section, the technical backgrounds upon which the proposed estimator is developed are reviewed first.

## 3. BACKGROUNDS OF DOMAIN TRANSFORMATION

We review the technical backgrounds of domain transformation in this section to facilitate the demonstration of the proposed method in Sections 4-6. The transformations of the porous domain described in Figure 3 necessitate the definition of design velocity on perturbed pore boundaries and material derivative of generic functions, which are, respectively, discussed in the following two sections.

3.1 Design velocity

Consider a smooth domain $(\Omega - \Omega_P^\eta)$ with a parameterized pore $(\Omega_P^\eta)$ with the size of $\eta$ located at an arbitrary point $\mathbf{X_c}$ in Figure 4(a). Let the pore be perturbed by an infinitely small amount $d\eta$, and the new domain in Figure 4(b) is denoted as $(\Omega - \Omega_P^{\eta-d\eta})$ with the perturbed pore $\Omega_P^{\eta-d\eta}$. It is noted that only one pore is considered in this transformation without porosity interactions.

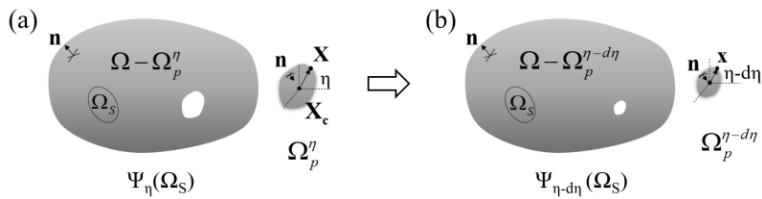

**Figure 4:** Domain transformation on the internal pore: (a) Parameterized domain, and (b) Perturbed domain.

Let the mapping in Figure 4 be smooth and invertible. We describe the mapping by a domain transformation where the original porous geometry in Figure 4(a) is referred to as a material domain with associated points denoted as **X**, and the perturbed geometry in Figure 4(b) is described by a spatial domain with points denoted as **x**.

We use a shape parameter $\eta$ to denote the amount of geometry change in the direction of perturbation, and express the shape transformation **T** between the two domains in Figure 4 by a linear and continuous transformation as:

$$\mathbf{T}(\eta):\left[\Omega-\Omega_P^{\eta}\right](\mathbf{X})\rightarrow\left[\Omega-\Omega_P^{\eta-d\eta}\right](\mathbf{x}) \tag{3.1}$$

For any point on pore surfaces, this transformation represents a scaling process where pore size is reduced by a small amount $d\eta$; whereas for any point *not* on pore surfaces, the mapping would not change its location:

$$\mathbf{T}(\eta):\mathbf{x}=(\mathbf{X}-\mathbf{X}_c)\eta+\mathbf{X}_c \qquad \mathbf{x}\in\Gamma_P \tag{3.2}$$

$$\mathbf{T}(\eta):\mathbf{x}=\mathbf{X} \qquad \mathbf{x}\notin\Gamma_P \tag{3.3}$$

Considering the amount of perturbation as pseudo-time, we define the design velocity [24] regarding the perturbations of pore surfaces in Equation (3.4). The design velocity on pore surfaces, specifically, is computed in Equation (3.5), while the velocity of points, which are not on pore surface, is assumed as zero.

$$\mathbf{V}\equiv\frac{d\mathbf{T}}{d\eta} \tag{3.4}$$

$$\mathbf{V}=\mathbf{X}-\mathbf{X}_c \qquad \mathbf{x}\in\Gamma_P \tag{3.5}$$

$$\mathbf{V}=\mathbf{0} \qquad \mathbf{x}\notin\Gamma_P \tag{3.6}$$

3.2 Material derivative

Given the design velocity in Equation (3.4), a material derivative ($\dot{\mathbf{z}}$) can be defined for the displacement field (**z**) on a perturbed point ($\mathbf{x}_\eta$) by:

$$\dot{\mathbf{z}}\equiv\frac{d}{d\eta}\mathbf{z}_\eta(\mathbf{x}_\eta)=\lim_{d\eta\to 0}\left[\frac{\mathbf{z}_\eta(\mathbf{x}+d\eta\mathbf{V}(\mathbf{x}))-\mathbf{z}(\mathbf{x})}{d\eta}\right]=\mathbf{z}'(\mathbf{x})+\nabla\mathbf{z}\mathbf{V}(\mathbf{x}) \tag{3.7}$$

where $d\eta$ represents an infinitesimal shape perturbation, the term $\nabla\mathbf{z}\mathbf{V}(\mathbf{x})$ is a convective term, and $\mathbf{z}'(\mathbf{x})$ is the spatial derivative representing the sensitivity of the displacement field **z** on the same spatial point **x** to the shape perturbations $d\eta$:

$$\mathbf{z}'=\lim_{d\eta\to 0}\left[\frac{\mathbf{z}_\eta(\mathbf{x})-\mathbf{z}(\mathbf{x})}{d\eta}\right] \tag{3.8}$$

To compute the shape sensitivity of an arbitrary function, we define a scalar domain functional ($\Psi$) in an integral form as:

$$\Psi=\iiint_{\Omega_\eta}g_\eta\left(\mathbf{z}_\eta\left(\mathbf{x}_\eta\right)\right)d\Omega \tag{3.9}$$

where $g_\eta(\mathbf{z}_\eta(\mathbf{x}_\eta))$ is a scalar function of the displacement field $\mathbf{z}_\eta(\mathbf{x}_\eta)$ defined over the perturbed domain $\Omega_\eta$. The material derivative of the domain functional is derived as [24]:

$$\Psi' \equiv \frac{d\Psi}{d\eta} = \iiint_\Omega g'(\mathbf{z}(\mathbf{x})) d\Omega + \iint_\Gamma g(\mathbf{z}(\mathbf{x})) V_n d\Gamma \qquad (3.10)$$

where $V_n$ is the normal component of the design speeds defined over the perturbed pore surfaces, and $g'(\mathbf{z}(\mathbf{x}))$ accounts for the material derivative of the displacement-dependent function to the shape perturbation. We point out that the term $g'(\mathbf{z}(\mathbf{x}))$ implicitly includes the derivative of displacements regarding shape perturbations. For demonstration purposes, let us consider a pointwise displacement functional defined at a node $\hat{\mathbf{x}}$ far from pore surfaces:

$$\Psi = \iiint_{\Omega_\eta} \delta(\mathbf{x}-\hat{\mathbf{x}}) z d\Omega \qquad (3.11)$$

where $z$ is the displacement component (scalar) along a user-defined direction. Based on Equation (3.10), its material derivative is:

$$\Psi' = \iiint_\Omega \delta(\mathbf{x}-\hat{\mathbf{x}}) z' d\Omega + \iint_\Gamma (\delta(\mathbf{x}-\hat{\mathbf{x}}) z) V_n(\hat{\mathbf{x}}) d\Gamma \qquad (3.12)$$

where since we assume the node $\hat{\mathbf{x}}$ does not locate on the pore surface, and as a result, the second term of Equation (3.12) can be dropped due to the zero design speed on the node ($V_n(\hat{\mathbf{x}}) = 0$), see Equation (3.6). The dependency of the displacement solution on the perturbation parameter $\eta$ is seen in the term $z'$.

## 4. SECOND-ORDER POROSITY SHAPE SENSITIVITY

Shape sensitivity describes the change of performance functions to infinitesimal shape variations. Specifically, to estimate the change of target functions from the original porous domain in Figure 3(a) to its perturbed counterpart in Figure 3(b), the shape sensitivity needs to be computed based on the pore surface perturbations, containing both the first-order and second-order sensitivity terms. While the low-order terms suffice to account for the transformation effects on individual pores, they ignore the pore-to-pore interactions along with higher-order shape variations. To this end, we review the first-order shape sensitivity in Appendix B and derive the second-order porosity sensitivity in this section.

4.1 Second-order functional derivative

We use $\tau_t$ and $\tau_s$ to represent two infinitesimal perturbations on the boundaries of two interactive pores in proximity in Figure 5. We associate the perturbations on the two pores with two design velocities $\mathbf{V}$ and $\mathbf{W}$ and define a two-parameter family of the perturbed domains ($\Omega_{ts}$) [34] via boundary transformations as:

$$\begin{cases} \tilde{\mathbf{x}} = \mathbf{x} + t\mathbf{V}(\mathbf{x}) + s\mathbf{W}(\mathbf{x}), & \tilde{\mathbf{x}} \in \partial\Omega_{ts}, \quad \mathbf{x} \in \partial\Omega_\eta \\ \Omega_{ts}: (t,s) \in [(-\tau_t,+\tau_t),(-\tau_s,+\tau_s)] \end{cases} \qquad (4.1)$$

where $t$ and $s$ are the two shape parameters associated with the two pores. It is clear that when $t=s=0$, the perturbed domain resembles the parameterized reference domain, i.e., $\Omega_{oo}=\Omega_\eta$.

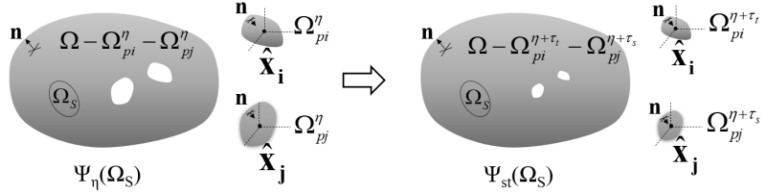

**Figure 5:** Domain transformations of two pores in proximity.

We define $U(\Omega_{ts})$ as a suitable functional space on the perturbed domain, and $U(\Omega_\eta)$ as a functional space on the parameterized domain $\Omega_\eta$. The weak form on the perturbed domain in Figure 5 can be derived by following Appendix B in Equation (4.2)-(4.4), and the performance function on the perturbed domain $\Omega_{ts}$ is written as Equation (4.5).

$$\tilde{a}(t,s;\tilde{\mathbf{z}},\tilde{\mathbf{v}}) = \tilde{l}(t,s;\tilde{\mathbf{v}}), \quad \forall \tilde{\mathbf{v}} \in U(\Omega_{ts}) \tag{4.2}$$

$$\tilde{a}(t,s;\tilde{\mathbf{z}},\tilde{\mathbf{v}}) = \iiint_{\Omega_{ts}} \boldsymbol{\varepsilon}(\tilde{\mathbf{v}}) : \boldsymbol{\sigma}(\tilde{\mathbf{z}}) d\Omega_{ts} \tag{4.3}$$

$$\tilde{l}(t,s;\tilde{\mathbf{v}}) = \iiint_{\Omega_{ts}} \tilde{\mathbf{v}}^{\mathbf{T}} \mathbf{f}^{\mathbf{b}} d\Omega_{ts} + \iint_{\Gamma_{ts}} \tilde{\mathbf{v}}^{\mathbf{T}} \mathbf{f}^{\mathbf{s}} d\Gamma_{ts} \tag{4.4}$$

$$\Psi(t,s;\tilde{\mathbf{z}}(t,s)) = \iiint_{\Omega_{ts}} g(\tilde{\mathbf{z}}(t,s)) d\Omega_{ts} \tag{4.5}$$

We also define a transformation tensor $\mathbf{F}$ to project the associated functional spaces between the reference domain and perturbed domain in Equation (4.6) with the properties of $\mathbf{F}$ defined in Equation (4.7).

$$\mathbf{F} \equiv \frac{d\tilde{\mathbf{x}}}{d\mathbf{x}} = \mathbf{I} + t\nabla\mathbf{V} + s\nabla\mathbf{W} \tag{4.6}$$

$$\begin{cases} \left.\frac{\partial}{\partial t}(\mathbf{F}^{-1})\right|_{t=s=0} = -\nabla\mathbf{V}, \quad \left.\frac{\partial}{\partial s}(\mathbf{F}^{-1})\right|_{t=s=0} = -\nabla\mathbf{W} \\ \left.\frac{\partial}{\partial t}(|\mathbf{F}|)\right|_{t=s=0} = \nabla\cdot\mathbf{V}, \quad \left.\frac{\partial}{\partial s}(|\mathbf{F}|)\right|_{t=s=0} = \nabla\cdot\mathbf{W} \end{cases} \tag{4.7}$$

We use the transformation tensor $\mathbf{F}$ to transform the functionals of Equation (4.2) from the perturbed functional space $U(\Omega_{ts})$ to the parameterized space $\Omega_\eta$ as in Equation (4.8)-(4.10).

$$a(t,s;\mathbf{z},\mathbf{v}) = l(t,s;\mathbf{v}), \quad \forall \mathbf{v} \in U(\Omega_\eta) \tag{4.8}$$

$$a(t,s;\mathbf{z},\mathbf{v}) = \frac{1}{4}\iiint_{\Omega_\eta} \left[\left(\mathbf{F}^{-\mathbf{T}}\nabla\mathbf{v} + \nabla\mathbf{v}^{\mathbf{T}}\mathbf{F}^{-1}\right) : \mathbb{C} : \left(\mathbf{F}^{-\mathbf{T}}\nabla\mathbf{z} + \nabla\mathbf{z}^{\mathbf{T}}\mathbf{F}^{-1}\right)|\mathbf{F}|\right] d\Omega_\eta \tag{4.9}$$

$$l(t,s;\mathbf{v}) = \iiint_{\Omega_\eta} \mathbf{v}^{\mathbf{T}} \mathbf{f}^{\mathbf{b}} |\mathbf{F}| d\Omega_\eta + \iint_{\Gamma_\eta} \mathbf{v}^{\mathbf{T}} \mathbf{f}^{\mathbf{s}} d\Gamma_\eta \tag{4.10}$$

We use the Gateaux derivatives [34] to define functional derivatives in the second-order sensitivity derivation, e.g., for an arbitrary function $g$:

$$\langle D_3 g(t,s;\mathbf{z}); \mathbf{v} \rangle \equiv \lim_{\alpha \to 0} \frac{g(t,s;\mathbf{z}+\alpha\mathbf{v}) - g(t,s;\mathbf{z})}{\alpha} \tag{4.11}$$

$$\left\langle D_{33}^2 g(t,s;\mathbf{z}):(\mathbf{v},\mathbf{w})\right\rangle \equiv \lim_{\beta \to 0} \frac{\left\langle D_3 g(t,s;\mathbf{z}+\beta\mathbf{w});\mathbf{v}\right\rangle - \left\langle D_3 g(t,s;\mathbf{z});\mathbf{v}\right\rangle}{\beta} \tag{4.12}$$

$$\left\langle D_{13}^2 g(t,s;\mathbf{z});\mathbf{v}\right\rangle \equiv \lim_{\beta \to 0} \frac{\left\langle D_1 g(t,s;\mathbf{z}+\beta\mathbf{v})\right\rangle - \left\langle D_1 g(t,s;\mathbf{z})\right\rangle}{\beta} \tag{4.13}$$

$$\left\langle D_1 g(t,s;\mathbf{z})\right\rangle \equiv \frac{\partial g}{\partial t} \tag{4.14}$$

where Equation (4.11) denotes the first-order derivative of the functional to its argument $\mathbf{z}$ in the direction of a function $\mathbf{v}$. Its subindex '3' indicates $\mathbf{z}$ is the third variable. Equation (4.12) is a second-order Gateaux derivative to its third argument $\mathbf{z}$ in the directions of functions $\mathbf{v}$ and $\mathbf{w}$, respectively. Similarly, Equation (4.13) defines a mixed second-order partial derivative for its first parameter t and a Gateaux derivative regarding its third argument $\mathbf{z}$ in the direction of $\mathbf{v}$. Equation (4.14) represents a regular first-order partial derivative to one of its shape parameters. To demonstrate the derivation of material derivative on performance functions to pore shape parameters, we select a pointwise displacement as the quantity of interest as:

$$\Psi(t,s;\tilde{\mathbf{z}}) = \iiint_{\Omega_{ts}} \delta(\mathbf{x}-\hat{\mathbf{x}})^T \tilde{\mathbf{z}} d\Omega_{ts} = \iiint_{\Omega_\eta} \delta(\mathbf{x}-\hat{\mathbf{x}})^T \mathbf{z}|\mathbf{F}|d\Omega_\eta = \Psi(t,s;\mathbf{z}) \tag{4.15}$$

4.2 Domain formulation of second-order porosity sensitivity

To derive the second-order sensitivity of the performance function in Equation (4.15) to two shape parameters ($t$ and $s$), we adopt Lagrangian multiplier method [35]. We construct a first-order Lagrangian ($L$) by summing the objective function of Equation (4.15) and the constraint in Equation (4.8) as:

$$L(t,s;\mathbf{z},\boldsymbol{\lambda}) = \Psi(t,s;\mathbf{z}) + a(t,s;\mathbf{z},\boldsymbol{\lambda}) - l(t,s;\boldsymbol{\lambda}) \tag{4.16}$$

where the Lagrangian $L(t,s;\mathbf{z},\boldsymbol{\lambda})$ is a function of shape parameters ($t$ and $s$), displacement variable $\mathbf{z}$, and the adjoint solution $\boldsymbol{\lambda}$. Vanishing the first derivatives of Equation (4.16) with respect to the two variables ($\mathbf{z}$ and $\boldsymbol{\lambda}$) results in the adjoint equation in Equation (4.17). Solving this adjoint equation results in the first-order porosity shape sensitivity in its domain formulation, see Appendix B.

$$\frac{\partial \Psi}{\partial t} = \frac{\partial L(t,s;\mathbf{z},\boldsymbol{\lambda})}{\partial t} \tag{4.17}$$

To compute the second-order shape sensitivity, we define a second-order Lagrangian as:

$$\begin{aligned}L_2(t,s;\mathbf{z},\boldsymbol{\lambda},\mathbf{P},\mathbf{Q}) \\ = \frac{\partial L(t,s;\mathbf{z},\boldsymbol{\lambda})}{\partial t} + \left[a(t,s;\mathbf{z},\mathbf{P}) - l(t,s;\mathbf{P})\right] + \left[\left\langle D_3 a(t,s;\mathbf{z},\boldsymbol{\lambda});\mathbf{Q}\right\rangle + \left\langle D_3 \Psi(t,s;\mathbf{z});\mathbf{Q}\right\rangle\right]\end{aligned} \tag{4.18}$$

where the second term on the right side represents the constraint of Equation (4.8) and the third term denotes the constraint of the adjoint equation (see Appendix B). The two constraints are associated with two Lagrangian multipliers $\mathbf{P}$ and $\mathbf{Q}$, respectively. Computing the second-order shape sensitivity, therefore, boils down to taking derivative of the second-order Lagrangian in Equation (4.18) in a smooth interval $s \in (-\tau_s, +\tau_s)$ in Equation (4.19). We provide the mathematical expression of the volume integration of the second-order derivative in Appendix C.

$$\frac{\partial^2 \Psi}{\partial t \partial s} = \frac{\partial L_2(t,s;\mathbf{z},\lambda,\mathbf{P},\mathbf{Q})}{\partial s} \tag{4.19}$$

4.3 Boundary formulation of second-order porosity sensitivity

In this section, the domain formulations of porosity sensitivities are converted to boundary formulations for ease of implementation. We transform the first-order sensitivity with the help of variational identities (see Appendix B) and obtain the boundary integral form as:

$$\Psi' = \frac{d(\Psi_\eta(\Omega_s))}{d\eta} = -\iint_{\Gamma_\eta^P} \left[\boldsymbol{\sigma}(\mathbf{z}_\eta):\boldsymbol{\varepsilon}(\lambda_\eta)\right] V_n^\eta d\Gamma \tag{4.20}$$

where $\Gamma_\eta^p$ refers to the pore surfaces parameterized by the shape parameter $\eta = t$, $V_n^\eta$ is the design speed along normal direction, $\boldsymbol{\sigma}(\mathbf{z}_\eta)$ and $\boldsymbol{\varepsilon}(\lambda_\eta)$ are the primary stress and adjoint strain fields on pore surfaces, respectively.

We use the Guillaume-Masmoudi lemma [34] to transform the second-order sensitivity. Following this lemma, the sensitivity's domain formulation can be transformed to its boundary equivalence as:

$$\langle D\Psi(t,s);\mathbf{V}\rangle|_{t=s=0} = \iiint_{\Omega_\eta} (\mathbf{G}:\nabla\mathbf{V} + \mathbf{h}\cdot\mathbf{V}) d\Omega_\eta = \iint_{\Gamma_\eta} (\mathbf{G}^\mathrm{T}\mathbf{V}\cdot\mathbf{n}) d\Gamma_\eta \tag{4.21}$$

where $\mathbf{G}$ is an arbitrary second-order tensor, and $\mathbf{h}$ is an arbitrary vector, respectively. With the help of the Guillaume-Masmoudi lemma, we transform the second-order porosity shape sensitivity into its boundary equivalence and drop all terms without design velocities as:

$$\begin{aligned}\langle D^2\Psi:(\mathbf{V},\mathbf{W})\rangle &= \langle D\langle D\Psi;\mathbf{V}\rangle;\mathbf{W}\rangle - \langle D\Psi;(\nabla\mathbf{V})\mathbf{W}\rangle \\ &= \iint_{\Gamma_\eta^P} \begin{bmatrix} \boldsymbol{\varepsilon}(\mathbf{P}_\eta):\boldsymbol{\sigma}(\mathbf{z}_\eta)\mathbf{I} + \boldsymbol{\varepsilon}(\mathbf{Q}_\eta):\boldsymbol{\sigma}(\lambda_\eta)\mathbf{I} \\ -\boldsymbol{\sigma}(\mathbf{P}_\eta)\nabla\mathbf{z}_\eta^\mathrm{T} - \boldsymbol{\sigma}(\mathbf{z}_\eta)\nabla\mathbf{P}_\eta^\mathrm{T} \\ -\boldsymbol{\sigma}(\mathbf{Q}_\eta)\nabla\lambda_\eta^\mathrm{T} - \boldsymbol{\sigma}(\lambda_\eta)\nabla\mathbf{Q}_\eta^\mathrm{T} \end{bmatrix} \mathbf{W}\cdot\mathbf{n} d\Gamma\end{aligned} \tag{4.22}$$

where $\mathbf{W}$ is the design speed vector on a parameterized pore surface $\Gamma_\eta^p$ with a normal direction $\mathbf{n}$, $\boldsymbol{\sigma}(\mathbf{z}_\eta)$, $\boldsymbol{\sigma}(\lambda_\eta)$, $\boldsymbol{\sigma}(\mathbf{P}_\eta)$ and $\boldsymbol{\sigma}(\mathbf{Q}_\eta)$ are the stress fields computed from the primary variables $\mathbf{z}_\eta$, adjoint variables $\lambda_\eta$, and two Lagrangian multipliers $\mathbf{P}_\eta$ and $\mathbf{Q}_\eta$, $\boldsymbol{\varepsilon}(\mathbf{P}_\eta)$ and $\boldsymbol{\varepsilon}(\mathbf{Q}_\eta)$ are the strain fields of the two multipliers, and $\nabla\mathbf{z}_\eta^T$, $\nabla\lambda_\eta^T$, $\nabla\mathbf{P}_\eta^T$ and $\nabla\mathbf{Q}_\eta^T$ are the spatial gradients. We point out that all the field variables ($\mathbf{z}_\eta$, $\lambda_\eta$, $\mathbf{P}_\eta$ and $\mathbf{Q}_\eta$) are computed on the parameterized pore surfaces $\Gamma_\eta^p$, such that not only these field variables but also the second-order sensitivity $\langle D^2\Psi:(\mathbf{V},\mathbf{W})\rangle$ are implicit functions of the pore shape parameter $\eta$.

Since we have demonstrated both the first and second-order porosity sensitivities are implicit functions of the pore parameter in Equation (4.20) and (4.22), we need to integrate the sensitivity fields to calculate the effects of accumulative shape changes from Figure 3(a) to Figure 3(d) in Section 5. To this end, we can directly account for the performance function of the porous domain in Figure 3(a) by the field variables computed on the reference domain in Figure 3(d).

## 5. POROUS DOMAIN APPROXIMATIONS

This section aims to quantify the difference of performance functions, defined on the original porous domain of Figure 3(a) and its reference domain in Figure 3(d), as a function of the pore shape parameter $\eta$. To this end, we can approximate a generic function value on the porous domain by using the shape parameter and the field solutions efficiently calculated on the reference domain. Our method starts with comparing the boundary value problems (BVPs) for the $\eta$-parameterized porous domains.

5.1 Pore shape-dependent BVP

We formulate the shape-dependent BVPs for the domain with fully-sized pores ($\eta=1$), for the porous domain with parameterized pores ($0<\eta<1$), and for the reference domain without pores ($\eta=0$) in their strong forms in Equations (5.1)-(5.3). We note that when the pore parameter increases to its largest ($\eta=1$), the parameterized BVP in Equation (5.2) resembles Equation (5.1), and when the parameter decreases to zero ($\eta=0$), the parameterized BVP in Equation (5.2) reduces to Equation (5.3).

$$\begin{cases} -\nabla \cdot \boldsymbol{\sigma}(\mathbf{z}) = \mathbf{f}^b & \mathbf{x} \in \Omega - \Omega_p \\ \mathbf{z} = \hat{\mathbf{z}} & \mathbf{x} \in \Gamma^h \\ \boldsymbol{\sigma}(\mathbf{z}) \cdot \mathbf{n} = \mathbf{f}^s & \mathbf{x} \in \Gamma^s \\ \boldsymbol{\sigma}(\mathbf{z}) \cdot \mathbf{n} = \mathbf{0} & \mathbf{x} \in \Gamma^P \end{cases} \quad (5.1)$$

$$\begin{cases} -\nabla \cdot \boldsymbol{\sigma}(\mathbf{z}_\eta) = \mathbf{f}^b & \mathbf{x} \in \Omega - \Omega_P^\eta \\ \mathbf{z}_\eta = \hat{\mathbf{z}} & \mathbf{x} \in \Gamma^h \\ \boldsymbol{\sigma}(\mathbf{z}_\eta) \cdot \mathbf{n} = \mathbf{f}^s & \mathbf{x} \in \Gamma^s \\ \boldsymbol{\sigma}(\mathbf{z}_\eta) \cdot \mathbf{n} = \mathbf{0} & \mathbf{x} \in \Gamma_P^\eta \end{cases} \quad (5.2)$$

$$\begin{cases} -\nabla \cdot \boldsymbol{\sigma}(\mathbf{z}) = \mathbf{f}^b & \mathbf{x} \in \Omega \\ \mathbf{z} = \hat{\mathbf{z}} & \mathbf{x} \in \Gamma^h \\ \boldsymbol{\sigma}(\mathbf{z}) \cdot \mathbf{n} = \mathbf{f}^s & \mathbf{x} \in \Gamma^s \end{cases} \quad (5.3)$$

Parameterization of the performance function in Equation (2.5) by using the pore shape parameter on the three different domains are defined in Equations (5.4)-(5.6).

$$\Psi_1 = \iiint_{\Omega_S} g(\mathbf{z}) d\Omega \qquad \Omega_S \subset \Omega - \Omega_p \quad (5.4)$$

$$\Psi_\eta = \iiint_{\Omega_S} g(\mathbf{z}_\eta) d\Omega \qquad \Omega_S \subset \Omega - \Omega_P^\eta \quad (5.5)$$

$$\Psi_0 = \iiint_{\Omega_S} g(\mathbf{z}) d\Omega \qquad \Omega_S \subset \Omega \quad (5.6)$$

To approximate the difference of the performance functions between the original porous domain ($\eta=1$) and the reference domain ($\eta=0$), we formulate their difference by invoking the accumulative shape sensitivities during the domain transformations from Figure 3(a) to (d):

$$\Psi_1 - \Psi_0 = \mathcal{D}_{0 \leq \eta \leq 1} \Psi \tag{5.7}$$

where the entire transformation ($\mathcal{D}_{0 \leq \eta \leq 1} \Psi$) contains two processes: a shape varying process ($\mathcal{D}_{\xi \leq \eta \leq 1} \Psi$) with pore sizes continuously decreasing in the interval ($\xi \leq \eta \leq 1$), and a topology changing process ($\mathcal{D}_{0 \leq \eta \leq \xi} \Psi$) with the decreased pores removed ($0 \leq \eta \leq \xi$, $\xi = 0^+$). Equation (5.7) is, therefore, expanded as:

$$\mathcal{D}_{0 \leq \eta \leq 1} \Psi = \mathcal{D}_{0 \leq \eta \leq \xi} \Psi + \mathcal{D}_{\xi \leq \eta \leq 1} \Psi \tag{5.8}$$

where the first term is approximated by the topological sensitivity in Equation (5.9), and the second term is further expanded by two terms in Equation (5.10).

$$\mathcal{D}_{0 \leq \eta \leq \xi} \Psi \equiv \mathcal{D}_{topo} = Vol(\xi) \mathcal{T}(\hat{\mathbf{x}}) \tag{5.9}$$

$$\mathcal{D}_{\xi \leq \eta \leq 1} \Psi \equiv \mathcal{D}_{shape} = \mathcal{D}_{shape}^1 + \mathcal{D}_{shape}^2 \tag{5.10}$$

where the term $\mathcal{D}_{shape}^1$ stands for the change of performance functions due to the accumulative first-order variations of pore shapes, and we can rewrite it in an integral form as Equation (5.11) by assuming the sensitivity field is continuously differentiable. In addition, $\mathcal{D}_{shape}^2$ represents the influence of the second-order shape variations on performance functions, and we can decompose $\mathcal{D}_{shape}^2$ to two terms in Equation (5.12): a second-order estimation term ($\mathcal{D}_{self}^2$) based on the shape change of each individual pore and an interactive term ($\mathcal{D}_{int}^2$) accounting for interactive effects between two pores. The two terms are further expanded as in Equations (5.13) and (5.14).

$$\mathcal{D}_{shape}^1 = \int_\xi^1 \left( \frac{d\Psi_\eta}{d\eta} \right) d\eta \tag{5.11}$$

$$\mathcal{D}_{shape}^2 = \mathcal{D}_{self}^2 + \mathcal{D}_{int}^2 \tag{5.12}$$

$$\mathcal{D}_{self}^2 = \frac{1}{2} \int_\xi^1 \left( \int_\xi^1 \left( \frac{d^2\Psi_\eta}{d\eta_i d\eta_i} \right) d\eta_i \right) d\eta_i + \frac{1}{2} \int_\xi^1 \left( \int_\xi^1 \left( \frac{d^2\Psi_\eta}{d\eta_j d\eta_j} \right) d\eta_j \right) d\eta_j, \quad i \neq j \tag{5.13}$$

$$\mathcal{D}_{int}^2 = \int_\xi^1 \left( \int_\xi^1 \left( \frac{d^2\Psi_\eta}{d\eta_i d\eta_j} \right) d\eta_i \right) d\eta_j, \quad i \neq j \tag{5.14}$$

where $i$ and $j$ denote the pore indices in an interactive pore-pair, the integrand in Equation (5.11) is the first-order shape sensitivity in Equation (4.20), and the integrands in Equations (5.13) and (5.14) are computed by the second-order shape sensitivities in Equation (4.22).

We have demonstrated the difference of performance functions on the original porous domain and the reference domain can be approximated by integrating the first and second-order sensitivities to the pore shape parameter $\eta$. However, as we point out in Equation (4.20) and (4.22) the sensitivities are implicit functions of $\eta$, performing analytical integration of the sensitivities is infeasible. To address this difficulty, we adopt exterior approximations in the next section to formulate the sensitivity fields as explicit functions of pore shape parameters, such that the integrations in Equations (5.11)-(5.14) can be performed analytically.

## 5.2 Exterior approximations

The aim of exterior formulations in this section is to approximate parameterized field variables on the perturbed pore surfaces as explicit functions of the shape parameter $\eta$ that facilitates the integration of sensitivity fields in the domain transformations in Figure 3(a)-(d).

We assume the field variables $\mathbf{z}_\eta$ and $\mathbf{z}_0$ satisfy the BVP for the parameterized domain and the reference domain in Equation (5.2) and (5.3), respectively, and we suppose there is a linear relationship between the two displacement fields by offsetting a residual field $\tilde{\mathbf{z}}_\eta$ as:

$$\mathbf{z}_\eta = \mathbf{z}_0 + \tilde{\mathbf{z}}_\eta \tag{5.15}$$

Based on linear elastic constitutive in Equations (5.16), we readily see the stress field of the offsetting displacement equals the difference between the stress fields of the parameterized and referenced fields in Equation (5.17).

$$\begin{cases} \boldsymbol{\sigma}_0 = \mathbb{C}:\boldsymbol{\varepsilon}_0 = \mathbb{C}:\left(\nabla \mathbf{z}_0 + \nabla \mathbf{z}_0^{\mathrm{T}}\right)/2 \\ \boldsymbol{\sigma}_\eta = \mathbb{C}:\boldsymbol{\varepsilon}_\eta = \mathbb{C}:\left(\nabla \mathbf{z}_\eta + \nabla \mathbf{z}_\eta^{\mathrm{T}}\right)/2 \end{cases} \tag{5.16}$$

$$\boldsymbol{\sigma}(\tilde{\mathbf{z}}_\eta) = \boldsymbol{\sigma}(\mathbf{z}_\eta - \mathbf{z}_0) = \boldsymbol{\sigma}(\mathbf{z}_\eta) - \boldsymbol{\sigma}(\mathbf{z}_0) \tag{5.17}$$

Subtracting Equation (5.3) from (5.2) and using the Equation (5.17), we obtain the BVP based on the residual field ($\tilde{\mathbf{z}}_\eta$) as:

$$\begin{cases} -\nabla \cdot \boldsymbol{\sigma}(\tilde{\mathbf{z}}_\eta) = \mathbf{0} & \mathbf{x} \in \Omega - \Omega_p^\eta \\ \tilde{\mathbf{z}}_\eta = \mathbf{0} & \mathbf{x} \in \Gamma^h \\ \boldsymbol{\sigma}(\tilde{\mathbf{z}}_\eta) \cdot \mathbf{n} = \mathbf{0} & \mathbf{x} \in \Gamma^s \\ \boldsymbol{\sigma}(\tilde{\mathbf{z}}_\eta) \cdot \mathbf{n} = -\boldsymbol{\sigma}_0 \cdot \mathbf{n} & \mathbf{x} \in \Gamma_p^\eta \end{cases} \tag{5.18}$$

Recall our assumptions in Equation (2.1), i.e., we assume pores are much smaller than domains and they are far away from domain surfaces. Under this assumption, the residual field $\tilde{\mathbf{z}}_\eta$ in Equation (5.18) can be approximated by $\tilde{\mathbf{z}}_\eta^*$ in an exterior Neumann formulation [36] on the parameterized domain as:

$$\begin{cases} -\nabla \cdot \boldsymbol{\sigma}(\tilde{\mathbf{z}}_\eta^*) = \mathbf{0} & \mathbf{x} \in R^n - \Omega_p^\eta \\ \tilde{\mathbf{z}}_\eta^* = \mathbf{0} & \mathbf{x} \to \infty \\ \boldsymbol{\sigma}(\tilde{\mathbf{z}}_\eta^*) \cdot \mathbf{n} = -\boldsymbol{\sigma}_0 \cdot \mathbf{n} & \mathbf{x} \in \Gamma_p^\eta \end{cases} \tag{5.19}$$

Since we are interested in the field solutions on the original porous domain, after several algebraic operations (see Appendix D), we reformulate the exterior BVP on the porous domain in Equation (5.20).

$$\begin{cases} -\nabla \cdot \boldsymbol{\sigma}(\mathbf{z}_E) = \mathbf{0} & \mathbf{X} \in R^n - \Omega_p^1 \\ \mathbf{z}_E = \mathbf{0} & \mathbf{X} \to \infty \\ \boldsymbol{\sigma}(\mathbf{z}_E) \cdot \mathbf{n} = -\boldsymbol{\sigma}_0 \cdot \mathbf{n} & \mathbf{X} \in \Gamma_p^1 \end{cases} \tag{5.20}$$

where since the exterior solution ($\mathbf{z}_E$) is defined over the original porous domain ($\mathbf{X}$), it is independent of the shape parameter $\eta$. Furthermore, the parameterized stress and strain fields (defined over the parameterized domain $\mathbf{x}$) can be explicitly expressed as the functions of the pore shape parameter $\eta$ in Equation (5.21). In a similar approach, the adjoints and Lagrangian multipliers on the parameterized domain can be approximated by their exterior solutions in Equations (5.22)-(5.24).

$$\begin{cases} \boldsymbol{\sigma}(\mathbf{z}_\eta(\mathbf{x})) = \boldsymbol{\sigma}(\mathbf{z}_0) + \boldsymbol{\sigma}(\mathbf{z}_E(\mathbf{X})) \\ \boldsymbol{\varepsilon}(\mathbf{z}_\eta(\mathbf{x})) = \boldsymbol{\varepsilon}(\mathbf{z}_0) + \boldsymbol{\varepsilon}(\mathbf{z}_E(\mathbf{X})) \end{cases} \tag{5.21}$$

$$\begin{cases} \boldsymbol{\sigma}(\boldsymbol{\lambda}_\eta(\mathbf{x})) = \boldsymbol{\sigma}(\boldsymbol{\lambda}_0) + \boldsymbol{\sigma}(\boldsymbol{\lambda}_E(\mathbf{X})) \\ \boldsymbol{\varepsilon}(\boldsymbol{\lambda}_\eta(\mathbf{x})) = \boldsymbol{\varepsilon}(\boldsymbol{\lambda}_0) + \boldsymbol{\varepsilon}(\boldsymbol{\lambda}_E(\mathbf{X})) \end{cases} \tag{5.22}$$

$$\begin{cases} \boldsymbol{\sigma}(\mathbf{Q}_\eta(\mathbf{x})) = \boldsymbol{\sigma}(\mathbf{Q}_E(\mathbf{X}))/\eta \\ \nabla \mathbf{Q}_\eta(\mathbf{x}) = \nabla \mathbf{Q}_E(\mathbf{X})/\eta \end{cases} \tag{5.23}$$

$$\begin{cases} \boldsymbol{\sigma}(\mathbf{P}_\eta(\mathbf{x})) = \boldsymbol{\sigma}(\mathbf{P}_E(\mathbf{X}))/\eta \\ \nabla \mathbf{P}_\eta(\mathbf{x}) = \nabla \mathbf{P}_E(\mathbf{X})/\eta \end{cases} \tag{5.24}$$

With the parameterized solutions expressed as functions of pore shape parameters in Equations (5.21)-(5.24), the sensitivity fields in Equation (4.20) and (4.22) can be explicitly integrated with $\eta$. Therefore, we can estimate the variation of performance functions during the domain transformation from Figure 3(a) to Figure 3(d) by the explicit integration of sensitivity fields in the next section.

## 6. SECOND-ORDER POROSITY ESTIMATOR

The purpose of this section is to formulate a second-order estimator to predict the influence of porosity on elastic performance functions where the proposed estimator combines the porosity sensitivity fields in Sections 4 and the integration formulations in Section 5.

Since our estimator aims to predict the performance function on a domain with full-size pores (see Figure 3(a)) by using field solutions from the reference domain in Figure 3(d), all variations during domain transformations from Figure 3(a) to Figure 3(d) need to be accounted for. To quantify the impacts of domain transformations on the performance function, we utilize pore shape ($\eta$) dependent sensitivity fields (see Section 4) to compute the change of functions in a parameterized domain regarding infinitesimal perturbations. With the help of exterior formulations, the sensitivity fields in Equations (4.20) and (4.22), which are implicitly based on $\eta$, are now converted to explicit functions of $\eta$ in Equations (5.21)-(5.24). We integrate the sensitivities to $\eta$ in the range of $0 \leq \eta \leq 1$, and demonstrate the integrations of the first-order shape sensitivity and the second-order shape sensitivity in Equation (6.1) and (6.2), respectively. Since we use topology sensitivity to estimate the quantity changes during the domain variation from Figure 3(c) to Figure 3(d), its contributions in Equation (6.3) are also included.

$$\mathcal{D}^1_{shape} = \int_{\Gamma^p} \left\{ -\frac{V_n}{2} [\boldsymbol{\sigma}_0(\mathbf{z}) + \boldsymbol{\sigma}_E(\mathbf{z})] : [\boldsymbol{\varepsilon}_0(\boldsymbol{\lambda}) + \boldsymbol{\varepsilon}_E(\boldsymbol{\lambda})](1 - \xi^2) \right\} d\Gamma \tag{6.1}$$

$$\mathcal{D}^2_{shape} = \iint_{\Gamma_r} \begin{bmatrix} [\boldsymbol{\sigma}_0(\mathbf{z}) + \boldsymbol{\sigma}_E(\mathbf{z})] : \boldsymbol{\varepsilon}_E(\mathbf{P})\mathbf{I} + [\boldsymbol{\sigma}_0(\boldsymbol{\lambda}) + \boldsymbol{\sigma}_E(\boldsymbol{\lambda})] : \boldsymbol{\varepsilon}_E(\mathbf{Q})\mathbf{I} \\ -\boldsymbol{\sigma}_E(\mathbf{P})[\nabla\mathbf{z}_0 + \nabla\mathbf{z}_E]^T - [\boldsymbol{\sigma}_0(\mathbf{z}) + \boldsymbol{\sigma}_E(\mathbf{z})]\nabla\mathbf{P}_E^T \\ -\boldsymbol{\sigma}_E(\mathbf{Q})[\nabla\boldsymbol{\lambda}_0 + \nabla\boldsymbol{\lambda}_E]^T - [\boldsymbol{\sigma}_0(\boldsymbol{\lambda}) + \boldsymbol{\sigma}_E(\boldsymbol{\lambda})]\nabla\mathbf{Q}_E^T \end{bmatrix} \mathbf{W} \cdot \mathbf{n}(1-\xi)^2 d\Gamma_p \qquad (6.2)$$

$$\mathcal{D}_{topo} = Vol(\xi)\mathcal{T}_{topo}(\hat{\mathbf{x}}) \qquad (6.3)$$

We note that while $\mathcal{D}^1_{shape}$ and $\mathcal{D}_{topo}$ represent the impacts of the first-order shape variation and topological change separately from each pore, $\mathcal{D}^2_{shape}$ accounts for not only the second-order shape variation on individual pores but also their interactions. For the benchmark example with only two pores in Figure 3, we can readily show that a generic elastic quantity of interests $\Psi(\Omega_S)$ on the porous domain can be estimated by its counterpart on the reference domain $\Psi_0(\Omega_S)$ and the added sensitivity fields:

$$\Psi(\Omega_S) = \Psi_0(\Omega_S) + \mathcal{D}_{pore} \qquad (6.4)$$

$$\mathcal{D}_{pore} = \mathcal{D}_{topo} + \mathcal{D}^1_{shape} + \mathcal{D}^2_{shape} \qquad (6.5)$$

Engineered metallic components, however, often contain many process-induced pores. We, therefore, extend the two-pore estimator in Equation (6.4) to the scenario of multiple pores by aggregating all pairwise estimations in Equation (6.6).

$$\Psi(\Omega_S) = \Psi_0(\Omega_S) + \sum_{1 \le i \le n}(\mathcal{D}_{topo})_i + \sum_{1 \le i \le n}(\mathcal{D}^1_{shape})_i + \sum_{1 \le i \ne j \le n}(\mathcal{D}^2_{shape})_{i,j} \qquad (6.6)$$

where $n$ is the total number of pores, and $i$ and $j$ denote two different pores in a pair. We point out the second-order estimator of Equation (6.6) can reduce to its first-order counterpart [28,31] by dropping the second-order terms as in Equation (6.7).

$$\Psi(\Omega_S) = \Psi_0(\Omega_S) + \sum_{1 \le i \le n}(\mathcal{D}_{topo})_i + \sum_{1 \le i \le n}(\mathcal{D}^1_{shape})_i \qquad (6.7)$$

We note that Equation (6.7) performs the approximation by simply adding independent topological and shape sensitivities on each pore, without considering their interactions. We demonstrate in numerical experiments (see Section 0) that the first-order estimator suffices for scenarios where pores locate far from each other but may fall short when pores are aggregated in certain regions, which is a common observation on manufacturing metallic parts. By contrast, the proposed estimator provides higher prediction accuracy by considering higher-order terms and pore-to-pore interactions.

In summary, we provide the overall framework of the proposed second-order porosity estimator as shown in Algorithm 1. A few points are noteworthy: (1) BVPs are never solved on the porous domain. In contrast, the field variables, including primary solutions, adjoint solutions, and Lagrangian multipliers, are computed on the reference domain (no pores) which significantly reduces meshing difficulty. (2) All sensitivity fields are only required on pore surfaces with nonzero design speeds, and they are computed in *parameterized* boundary formulations which avoid the generation of complex volume elements and prevent meshing singularity. (3) The quantity change ($\Delta\Psi$) during domain transformation (pore shape perturbation) is estimated by the integration of sensitivity fields. While the sensitivities implicitly depend on the pore shape parameter ($\eta$), the usage of exterior solution is to reformulate sensitivities as explicit functions of $\eta$ such that the integration to $\eta$ can be performed analytically.

**Algorithm 1** Framework of the second-order porosity estimator for structural elastic behaviors
---
1: **Procedure** estimation of the elastic performance function ($\Psi$) on a porous domain
2: ▷ **Step-1**: material preparation and model initialization
3:    Generate tomography reconstruction of porosity distributions in a manufactured metallic component
4:    Reconstruct pores' morphology from either synthetic models or actual geometries from tomography
5:    Create the numerical model of the component with manufacturing-induced pores
6:    Create a reference numerical model by removing all pores
7: ▷ **Step-2**: computation of the field variables on the *reference* model
8:    Calculate the value of the performance function ($\Psi_0$) in Equation (5.6)
9:    Compute the primary solutions ($\mathbf{z}$) from Equation (5.3)
10:    Compute the adjoint solutions ($\boldsymbol{\lambda}$), and the two Lagrangian multipliers ($\boldsymbol{P}$ and $\boldsymbol{Q}$), see Appendix B and C
11: ▷ **Step-3**: calculation of sensitivity fields
12:    Compute the topological sensitivity, see Appendix A
13:    Compute the first and the second-order shape sensitivities in Equation (4.20) and (4.22), respectively
14: ▷ **Step-4**: Integration of sensitivity fields
15:    Quantify the change of performance function ($\Delta\Psi$) as the function of domain transformation
16:    Approximate the domain transformation via the integration of combined sensitivity fields
17:    Reformulate sensitivities as *explicit* functions of pore parameter ($\eta$) by exterior formulation in Section 5.2
18:    Integrate the approximated sensitivity fields w.r.t $\eta$ in Equations (6.1)-(6.3)
19: ▷ **Step-5**: Assembly of estimators
20:    Consider the integrated sensitivity field as estimators ($\mathcal{D}_{topo}$, $\mathcal{D}^1_{shape}$ and $\mathcal{D}^2_{shape}$)
21:    Sum the topological, the first-order, and second-order estimators in Equation (6.5) as $\mathcal{D}_{pore}$
22:    Estimate the performance function on the porous domain ($\Psi$) by Equation (6.6)
23: **End Procedure**

Even though our estimator utilizes the second-order shape sensitivity as the previous method in [27], the two approaches bear three major differences as illustrated in Figure 6 where the two methods are compared in an arbitrary rectangle domain with Dirichlet and Neumann boundary conditions on $\Gamma^h$ and $\Gamma^s$, respectively, and a generic target function in $\Omega_s$.

First, the previous method aims to estimate the variation of target functions when two close interactive *boundary* features are suppressed. It aims to simplify computer-aided design (CAD) models with surface features where internal CAD features are rare. In other words, it only applies to boundary features where the domain's topology does not change after defeaturing. In contrast, our method focuses on the defeaturing estimation of multiple interactive *internal* pores. Shape sensitivity, in our method, only approximates the effects of shape transformation between full-scale pores and their infinitesimal counterparts (small circle with dash lines in Figure 6(b)), while pore removals are approximated by topological sensitivity.

Second, the previous method assumes shape sensitivity fields as constants during shape perturbations, and as a result, the sensitivities are independent of shape parameters. Such assumptions are appropriate for small perturbations but may fall short in large shape variations where field solutions implicitly depend on shape parameters. In comparison, our estimator uses exterior formulations to express shape sensitivities as explicit functions of shape parameters, such that the accumulative effects of (continuous) shape perturbations can be accounted for by integrations.

Third, the boundary features in the previous work have *regular* shapes, e.g., $\Gamma^{wi}$ and $\Gamma^{wj}$ in Figure 6(a), since they are represented in Bézier form, which helps to simplify the associated deformation speeds but may have difficulty in extending to arbitrary geometries due to the lack of speed definitions. Our method allows *complex* pore geometries by expressing pore surfaces (e.g., $\Gamma^{pi}$ and $\Gamma^{pj}$ in Figure 6(b)) directly as the collections of nodal coordinates, and the associated design speeds are defined with respect to pore centroids.

In summary, the previous work in [27] differs from our approach in several aspects, and it is, therefore, not suitable for our studied material systems that contains a large number of interactive internal pores with complex morphologies, see Figure 1(c) and Figure 6. The efficacy of our method is demonstrated by various numerical experiments in the next section.

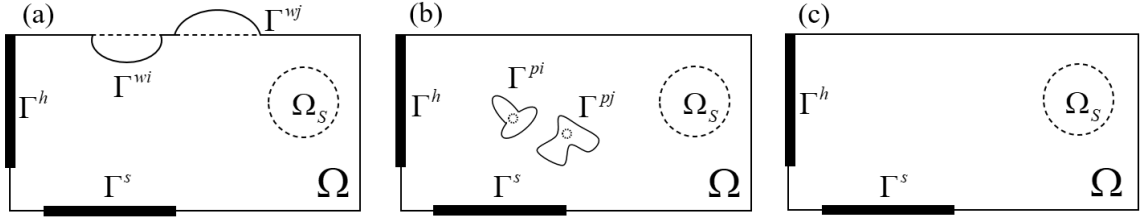

**Figure 6** Demonstration of similarities and differences between our method and the previous work in [27]: (a) Previous work studies multiple, interactive, and regularly-shaped boundary features (e.g., $\Gamma^{wi}$ and $\Gamma^{wj}$), (b) Our work focuses on multiple, interactive, internal features with complex morphologies (e.g., $\Gamma^{pi}$ and $\Gamma^{pj}$), and (c) Both methods require field solutions on the defeatured domain.

## 7. NUMERICAL EXPERIMENTS

In this section, we demonstrate the proposed method through several numerical experiments. The proposed porosity estimator is implemented in PYTHON scripts for both 2D and 3D cases. The elastic problems in Equation (2.6) and (2.7) are solved by ABAQUS [37], a commercial FE package. The exterior approximations are solved by the boundary element method [38]. All experiments were conducted on a 64-bit WINDOWS 10 machine with the following hardware: Intel I5-8250U CPU 4 cores running at 1.6 GHz with 16 GB of installed physical memory (RAM).

In all experiments, we neglect any metal polycrystalline microstructures (e.g., grain boundary, triple, or quad points) except for pores and we assume that materials are isotropic with perfect linear elastic properties (Young's modulus E=6.89e10 N/m$^2$, and Poisson's ratio $\upsilon$=0.35). The first example (Section 8.1) involves a 2D benchmark study where several pore parameters are studied for their influence on the accuracy of the proposed estimator. The second example (Section 8.2) is a case study on a 2D bracket with simulated pore spatial distributions. The third example (Section 8.3) involves a 3D hook model where the geometries of pores are represented by synthetic models with concave and convex radii. The last example (Section 8.4) is on a real casting component with tomography reconstructed porosity characteristics. In each example, we use the proposed second-order estimator to predict elastic quantities of interests and compare it against two other estimators as well as solutions from direct FEA:

1) *Direct FEA*: a porous model is created first. The BVPs in Equation (2.2) are then solved. The value of the performance function is denoted as $\Psi$. This approach is considered as the ground truth when compared with estimators. However, it is computationally expensive and hence should be avoided in practice.

2) *Topological sensitivity estimator (TSE)*: field variables are solved on a reference domain where its quantity of interests is denoted as $\Psi_0$. We represent each pore by an equivalent-sized circle void in 2D (or a sphere in 3D) and approximate the change of performance function by multiplying topological sensitivity with the pore's equivalent area (or volume in 3D), denoted as $\mathcal{D}_{topo}$ in Equation (A.6). As shown in experiments, this approach is inaccurate and only used here for comparison.

3) *First-order porosity sensitivity (FOE)*: the estimation on the change of performance function ($\mathcal{D}^1_{pore}$) is computed in Equation (6.7) which considers pore characteristics (morphology and distribution) but neglects pore-to-pore interactions.

4) *Second-order porosity estimator (SOE)*: by combining the first-order with the second-order shape sensitivities, the variation of target functions (i.e., defeaturing error) can be computed through Equation (6.6) and is denoted as $\mathcal{D}^2_{pore}$. The accuracy of this estimator is proved higher than the others in scenarios of closely clustered pores.

To quantify the accuracy of different estimators, an effectivity index [26] is defined as the ratio between the predicted function variation (i.e., defeaturing error) and the exact value as:

$$I_D = \frac{\mathcal{D}}{\Psi - \Psi_0} \tag{7.1}$$

where $\mathcal{D}$ can be $\mathcal{D}_{topo}$, $\mathcal{D}^1_{pore}$, or $\mathcal{D}^2_{pore}$. We note that the closer the value of $\mathcal{D}$ is to 1.0, the more accurate an estimator is. In the field of error estimation, for the quantities measured in the global norm, if the effectivity indices are in the range between 0.5 and 2.0, it would be considered acceptable [26]. However, for a local quantity (e.g., pointwise displacement), the effectivity index for a good estimator can be relaxed up to 10.0, as it is generally more difficult and expensive to obtain [39].

## 7.1 Benchmark study

In this section, a 2D cantilever beam is used as a benchmark example to study the influences of various pore parameters on the proposed estimator.

### *7.1.1 Pore distance*

In this example, we vary the distances between two pores to study the impact of distances on inter-pore interactions. As shown in Figure 7(a), the study is performed on a 2D cantilever beam with two identical circular pores. The dimension of the beam is 200 mm long and 100 mm wide. It is clamped on the left edge and a vertical tip load with a magnitude of 1000 N is applied on its up-right corner. The quantity of interest in this study is the vertical displacement at the bottom right corner of the beam, and its defeaturing error can be defined via Equation (6.6). While the radii of both pores are prescribed as 5 mm, their horizontal distance is varied from 1mm to 45 mm. It is noteworthy pore sizes in benchmarks do not necessarily indicate the sizes of actual pores on manufacturing components. Their relative sizes compared with the beam are only used for the parameter study.

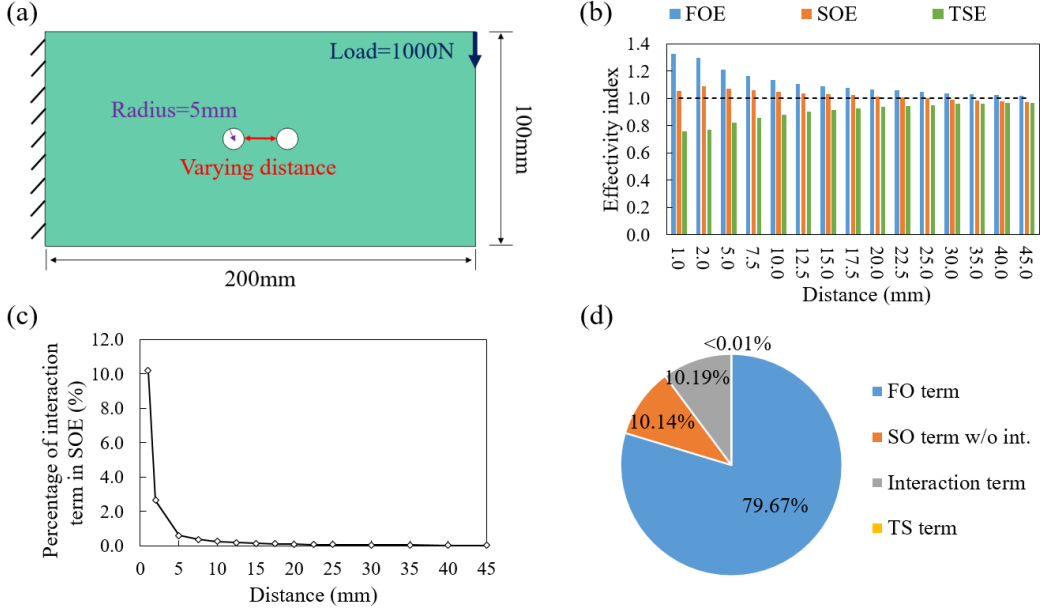

**Figure 7** Influence of inter-pore distances on estimation accuracy: (a) Dimensions and boundary conditions of the 2D cantilever beam with two closely distanced pores, (b) Effectivity indices of different estimators, (c) The value of interaction term with respect to the inter-pore distances, and (d) The percentages of different terms in the proposed SOE for the inter-pore distance as 1 mm.

We compared the accuracies of three different estimators, i.e., $\mathcal{D}_{topo}$, $\mathcal{D}^1_{pore}$, and the proposed $\mathcal{D}^2_{pore}$, to different pore-to-pore distances in Figure 7(b). It is evident when the distance is small (e.g., when distance is smaller than 5 mm), both $\mathcal{D}_{topo}$ and $\mathcal{D}^1_{pore}$ provide lower accuracy than $\mathcal{D}^2_{pore}$ (we emphasize that the closer an effectivity index is to 1.0, the more accurate the estimator is). On the other hand, as the distance becomes larger (e.g., when distance is larger than 20 mm), all three estimators result in similarly good predictions in this example.

The observation, that the accuracies of $\mathcal{D}_{topo}$ and $\mathcal{D}^1_{pore}$ start to deteriorate as pores are in closer proximity, can be explained by the following reason. When the distance decreases, dramatically increased porosity interactions cannot be accounted for by the two estimators. To verify this, the relation between pore-to-pore distances and values of the interaction term in Equation (5.14) is plotted in Figure 7(c). We notice a qualitative relationship between the interaction and distances from Figure 7(c). For example, when the distance decreases to 1 mm, the interaction term accounts for about 10% of the proposed estimator. But as the distance increases, the interaction effect gradually drops to a neglectable level.

To have a better understanding of the contributions from each term in the proposed second-order estimator, the percentage of each term in Equation (6.6) are compared in Figure 7(d) where the pore-to-pore distance is 1mm. It is observed the first-order (FO) terms in Equation (6.1) account for 79.7%, and the second-order (SO) terms together in Equation (6.2) account for about 20.3%. Specifically, the interaction term in Equation (5.14) accounts for 10.2%. We note that the value from the topological sensitivity (TS) term in Equation (6.3) is trivial (smaller than 0.01%) compared with others due to a small $\xi$ value we choose. Topological sensitivity is only used here to approximate a structural topology change where one (isolated and theoretically infinitesimally) small hole is introduced at each of the two pore locations. In a heuristic manner, we choose $\xi$ as 1% of each pore's actual size. It should also be noted that with the change of inter-pore distances, the values of terms in Figure 7(d) are subject to change.

From this experiment, we observe that the (first-order) TSE demonstrates a high-fidelity estimation in scenarios when two pores are at a far distance (weak pore-to-pore interactions), but its accuracy rapidly deteriorates when inter-pore distance decreases (strong pore-to-pore interactions). The second-order topological sensitivity, which captures the second-order impact of inserting an infinitesimally small *spherical* hole within a domain on various target functions [40], is arguably more accurate than its first-order counterpart. On a perturbed domain, the target function ($\Psi_\xi$) in a fixed region ($\Omega_S$) is written in the form of topological asymptotic expansion as:

$$\Psi_\xi(\Omega_S) = \Psi_0(\Omega_S) + g_1(\xi)\mathcal{T}^1_{topo} + g_2(\xi)\mathcal{T}^2_{topo} + R \tag{7.2}$$

where $\Psi_0$ is the target function in the same region $\Omega_S$, $T^1_{topo}$ and $T^2_{topo}$ are the first and second-order topological sensitivity fields, respectively, $g_1$ and $g_2$ are two positive monotonic functions depending on the size of pore $\xi$ such that $g_1 \to 0, g_2 \to 0$ as $\xi \to 0$, and the function $R$ represents higher-order terms. Divide Equation (7.2) by $g_2$, take the limit $\xi \to 0$, and as a result, we obtain the second-order topological derivative in Equation (7.3) which is simplified to Equation (7.4) with the definition of the first-order topological derivative (see Appendix A).

$$\mathcal{T}^2_{topo} = \lim_{\xi \to 0} \frac{\Psi_\xi(\Omega_S) - \Psi_0(\Omega_S) - g_1(\xi)\mathcal{T}^1_{topo}}{g_2(\xi)} \tag{7.3}$$

$$\mathcal{T}^2_{topo} = \lim_{\xi \to 0} \frac{\frac{d}{d\xi}\Psi(\Omega_S) - g'_1(\xi)\mathcal{T}^1_{topo}}{g'_2(\xi)} \tag{7.4}$$

where $g'_1$ and $g'_2$ are the derivatives of $g_1$ and $g_2$ to $\xi$, respectively. It is evident that deriving the exact expression of the second-order topological sensitivity (for our target function of the pointwise displacement in an elasto-static problem) in Equation (7.4) is non-trivial since it is a function of many variables, including its first-order counterpart $T^1_{topo}$, the characteristic of the target function $\Psi$, the size of the pore $\xi$, the domain's field solutions (e.g., primary and adjoint solutions), the selected forms of $g_1$ and $g_2$, the boundary conditions on the pore surfaces during transformation, etc. However, we can still obtain an insight into the second-order TSE from [41] which solves a simple Poisson's problem. It is reported the second-order TSE is only superior to its first-order counterpart when the pore size is significant (e.g., larger than 30% of the domain), but it does not provide significant accuracy improvement when pores are small (e.g., smaller than 10% of the domain). Considering our assumption in Equation (2.1) that our studied pores are process-induced pores with considerably small sizes ($\|\Omega_p\| \ll \|\Omega\|$) and the fact that the pore size only accounts for 3.5% of the domain size in this experiment, we believe the second-order TSE would not outperform the first-order TSE (if any) in this work. In addition, the second-order TSE ignores the porosity interactions and lacks the mechanism to estimate pores of complex shapes (rather than spheres), which are the major contributions of our estimators and will be further demonstrated in the following experiments.

*7.1.2 Pore size*

We study the impact of pore sizes on the proposed estimator in this benchmark example on a model in Figure 8(a). This example uses the same 2D cantilever beam as the previous study. In this example, the pore-to-pore distance is fixed as 1 mm but the radii of the two identical pores are varied from 1.5 mm to 10 mm. Again, the performance function is the vertical displacement at the right bottom corner.

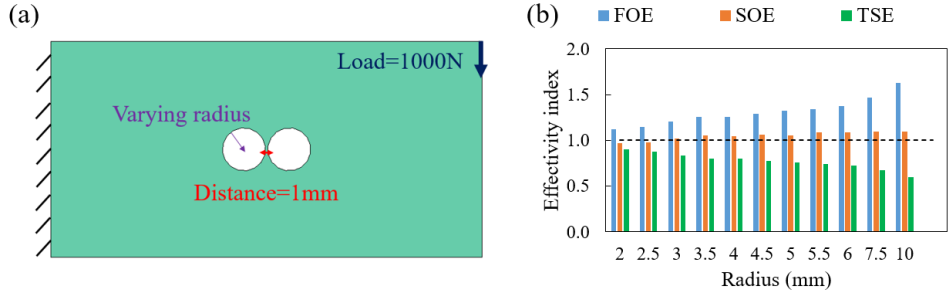

**Figure 8** Influence of pore sizes on estimation accuracy: (a) Boundary conditions of the cantilever beam containing two pores with the fixed distance, and (b) Effectivity indices of different estimators.

The accuracies of the three estimators are compared regarding different pore sizes in Figure 8(b). In this study, since the inter-pore distance is held constant, the larger pore sizes, the smaller ratio between inter-pore distances and pore sizes. We note smaller ratios, intuitively, represent stronger pore-to-pore interactions. As in Figure 8(b), when the pore sizes increase, the accuracies of both $\mathcal{D}_{topo}$ and $\mathcal{D}^1_{pore}$ start to deteriorate while $\mathcal{D}^2_{pore}$ is still maintained at high fidelity. For example, when the pore radii are 10mm, the effectivity index of $\mathcal{D}^2_{pore}$ is 1.1 while the indices of $\mathcal{D}_{topo}$ and $\mathcal{D}^1_{pore}$ are 1.6 and 0.6, respectively. The deteriorating accuracy is because strong porosity interactions are neglected by the two estimators.

### 7.1.3 Pore shape

Pore shape is an important factor in determining structural behaviors in many aspects. In this benchmark, we simplify pore shape as an ellipse. Its major and minor axis lengths are varied to study the influences on the accuracy of the proposed estimator. We use the same 2D cantilever beam as previous studies, but place two identical ellipse pores in the center and distanced by 1 mm as in Figure 9(a). We fix the ellipse's major axis as 5 mm and vary its minor axis between 3.5 mm and 6.5 mm.

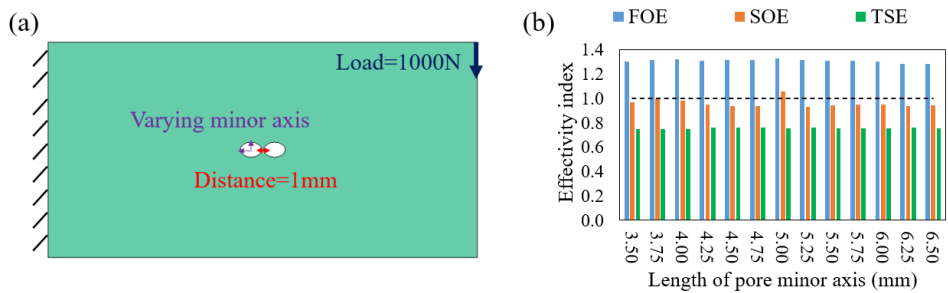

**Figure 9** Influence of pore shapes on estimation accuracy: (a) Boundary conditions of the cantilever beam containing two pores with varying aspect ratios, and (b) Effectivity indices of different estimators.

The quantity of interest is the vertical displacement at the beam's right bottom corner, and we compare the values of different estimators in Figure 9(b). Since the two pores are in proximity, the proposed second-order estimator outperforms the others due to strong interactions among pores.

### 7.1.4 Multiple pores

Manufactured metal components often contain many closely clustered pores. In the scenario of multiple pores, porosity interaction should account for all neighboring pores. Therefore, in the last benchmark example, we study the same beam model but place four pores in the close distance as shown in Figure 10(a).

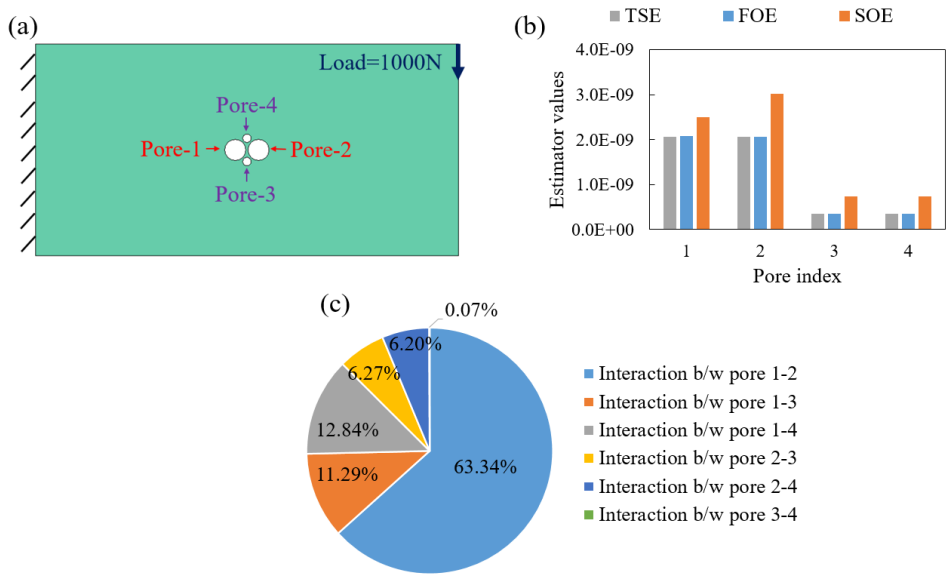

**Figure 10** Influence of multiple interactive pores on estimation accuracy: (a) Boundary conditions of the cantilever beam containing four neighboring pores with different sizes, (b) Comparison of different estimators for each pore, and (c) Comparisons of the pairwise interactions among the four pores.

In Figure 10(a), four circular pores are symmetrically placed in the center of the beam such that the (surface-to-surface) distances between the two large pores (i.e., between pore-1 and pore-2) are 1.0 mm, and the distance between large and small pores (e.g., between pore-1 and pore-3) is 0.78 mm, and the distance between the two small pores (i.e., between pore-3 and pore-4) is 7.0 mm. Radii of pore-1 and pore-2 are 5 mm and the radii of pore-3 and pore-4 are 2 mm, respectively. Similar to previous studies, the quantity of interest in this example is the vertical displacement at the bottom right corner of the beam. The defeaturing error of the pointwise displacement from the three estimators is compared with FEA results in Table 2. By comparing the effectivity indices, we find the proposed second-order estimator ($I_{pore}^2$) provides the best approximation due to its accountability for inter-pore interactions.

**Table 2:** Comparison of defeaturing errors between different estimators.

| FEA | $\mathcal{D}_{topo}$ | $I_{topo}$ | $\mathcal{D}_{pore}^1$ | $I_{pore}^1$ | $\mathcal{D}_{pore}^2$ | $I_{pore}^2$ |
|---|---|---|---|---|---|---|
| 6.21E-9 | 4.81E-9 | 1.29 | 4.82E-9 | 1.29 | 6.89E-9 | 0.91 |

To further illustrate the reason that the second-order estimator is superior to the others, the three estimators are compared on each pore in Figure 10(b). Investigation of this figure provides us with two observations. First, by considering the difference between the first and second-order estimators, it is obvious the interactions are significant in all pores. Second, by comparing the interaction with its first-order estimator, it is obvious the effects of interactions are more significant for small pores. When multiple pores are clustered together, accounting for all interactions between different pairs can be nontrivial. For example, in this example which only has four pores, we have six pairwise interactions. To understand their relative importance, their values are compared in Figure 10(c). We point out the interactions between the two large pores, (i.e., between pore-1 and pore-2) are dominant, accounting for 63% of total interaction effects. Interactions between large and small pores (e.g., between pore-1 and pore-3) account for 6%-13%. On the other hand, the interaction between the two small pores (i.e., between pore-3 and pore-4) appears trivial (about 0.07%) due to their small sizes and long distances.

Through the benchmark studies, we have demonstrated the proposed second-order estimator is more robust and accurate than the others under the investigation of different pore distances, sizes, shapes, and numbers. In real applications, it is common many pores are clustered together in proximity. Under such a scenario, it would be laborious to compute all pairwise interactions. Since we have shown interactions between closely distanced large pores dominate others, in the following case studies, we only focus on the nearest-neighbor interactions between large pores.

7.2 Case study: 2D bracket

Our first case study aims to investigate the impact of porosity on a 2D bracket whose dimensions and boundary conditions are shown in Figure 11(a). While the bracket is constrained by the fixture hole on the right side, two vertical loads are applied from the top and bottom sides on the rectangular arms on the left side, see the boundary conditions in Figure 11(b). The quantity of interest is the relative narrowing of the gap between the rectangular arms.

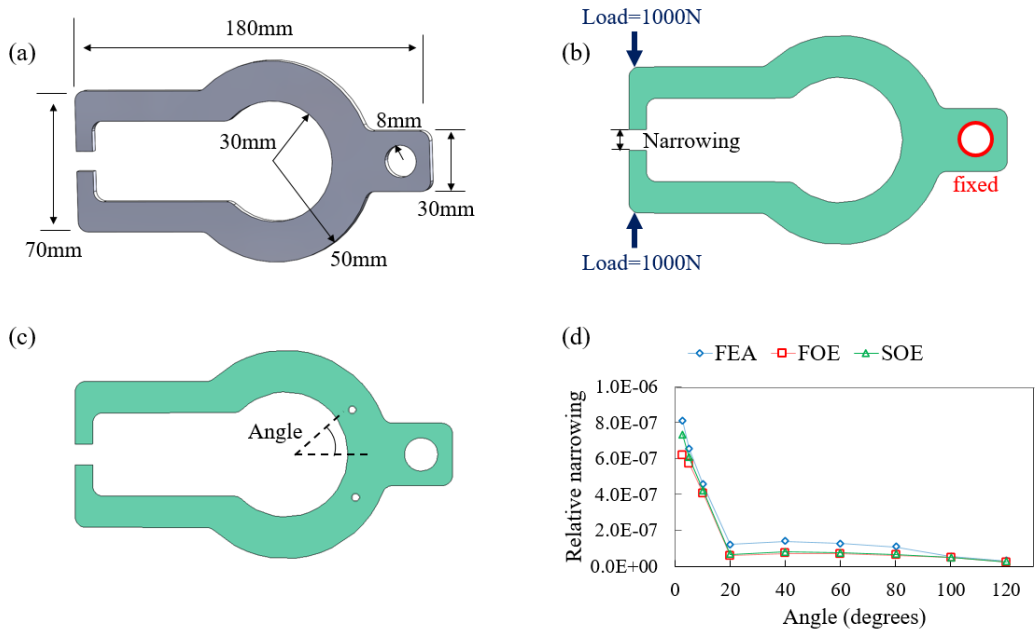

**Figure 11** Cast study on a 2D bracket: (a) Dimensions, (b) Boundary conditions on the 2D plane stress representation, (c) The bracket model with two symmetric circular pores, and (d) Predicted gap narrowing to the pores' angular positions.

We first study the relative narrowing as a function of the angular positions of two symmetric circular pores as shown in Figure 11(c). Both pores have the same diameters of 3 mm. When the angle varies from 2.5 degrees to 120 degrees, the defeaturing errors of the gap narrowing are demonstrated in Figure 11(d). It is clear that when the angles are small, the pore distances are also small, resulting in strong inter-pore interactions. For example, when the angle is 2.5 degrees, the distance is 3.8 mm and when the angle is 120 degrees, the distance increases to 70 mm. Thus, when the angle is small, the two pores are at a close distance and the second-order estimator provides a closer estimation to FEA. On the other hand, as the angle increases, the interaction becomes weaker where we observe similar predictions from different estimators.

To test the proposed estimator with manufacturing-induced porosity, we simulate the porosity spatial distributions via casting simulation software, MAGMASOFT [42]. As shown in Figure 12(a), the simulation result indicates that casting pores would cluster in the central region between the fixture hole and its inner surface. To simulate the impacts of pore's actual morphologies, six pores of distinct morphologies, which are reconstructed from the CT images from [18], are placed into the region as shown in Figure 12(b). Due to their relative positions, the pores are grouped with their nearest neighbors into three pairs. The average sizes and distances in each pair are summarized in Table 3. We assume pores only interact with their nearest neighbors, and the interactions between different pairs are neglected due to long distances.

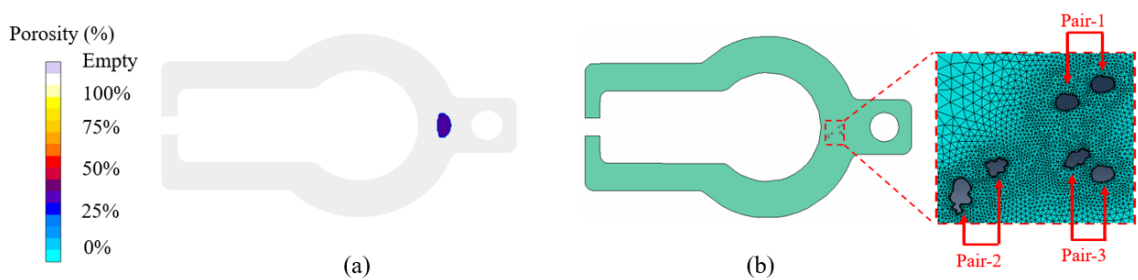

**Figure 12:** (a) Top view of the porosity spatial distributions modeled by MAGMASOFT, and (b) Spatial distributions and morphologies of the simulated casting pores.

The defeaturing error of the gap narrowing from FEA is 2.19e-7 mm, and the predictions from the first- and second-order estimators are 1.22e-7 mm and 1.87e-7 mm, respectively. It is therefore shown the second-order estimator (with the effectivity index of 1.17) outperforms its first-order counterpart (with the effectivity index of 1.79). To further compare the two estimators, their values in each pair are summarized in Table 3 where the sum of the fourth column ($\mathcal{D}^1_{pore}$) equals the total value of the first-order estimation and the sum of the fifth columns ($\mathcal{D}^2_{pore}$) equals the second-order estimator. The percentages of the interaction effects of different pairs range from 2.45% to 8.81% in the seventh column. It should be noted that the percentage values are specific to each pair. The values are dependent on pore locations, sizes, distances, shapes, and values of local field variables (e.g., displacement and stress).

**Table 3:** Values of parameters and porosity estimators for the 2D bracket model in Figure 12.

|        | Ave. diameter (mm) | Distance (mm) | $\mathcal{D}^1_{pore}$ | $\mathcal{D}^2_{pore}$ | $\mathcal{D}^2_{int}$ | $\mathcal{D}^2_{int}/\mathcal{D}^2_{pore}$ |
|--------|-------------------|---------------|------------------------|------------------------|-----------------------|--------------------------------------------|
| Pair-1 | 1.00              | 2.24          | 1.39E-8                | 1.50E-8                | 6.30E-10              | 4.19%                                      |
| Pair-2 | 1.28              | 2.50          | 9.01E-8                | 1.46E-7                | 1.29E-8               | 8.81%                                      |
| Pair-3 | 1.10              | 1.81          | 1.81E-8                | 2.52E-8                | 6.17E-10              | 2.45%                                      |

7.3 Case study: 3D hook

In this case study, the proposed estimator is applied to a 3D hook model whose dimensions and boundary conditions are illustrated in Figure 13(a). The model is subject to a vertical load inside the loading hole and the quantity of interest is a pointwise vertical displacement on the hook's shoulder section. 15 pore models of different sizes are placed inside the hook model as shown in Figure 13(b). For each pore, its distance with every neighbor is computed and we assume only the nearest neighbor is considered for pore interactions. Also, if the distance is more than five times the pore sizes, we assume the interaction is too trivial for consideration. As a result, there are 10 out of the 15 pores, which have interactive neighbors, are labeled for interaction in Figure 13(b). For the labeled pores, a summary of their sizes, interacting neighbors, and inter-pore distances are provided in Table 4. It is noted from this table we only consider the one-to-one pore interactions for the sake of simplification. For example, in the hook's neck region where pore-3, pore-4, pore-7, and pore-8 are located closely, we only consider the interactions in the pair 3-4 and the pair 7-8. To simulate porosity morphology more realistically, a 3D synthetic pore model [13] is utilized in this section. This model, which has better pore presentations than ellipsoid and sphere shapes, is generated by intersecting three mutually perpendicular equivalent ellipsoids at their geometric centers as shown in Figure 13(c).

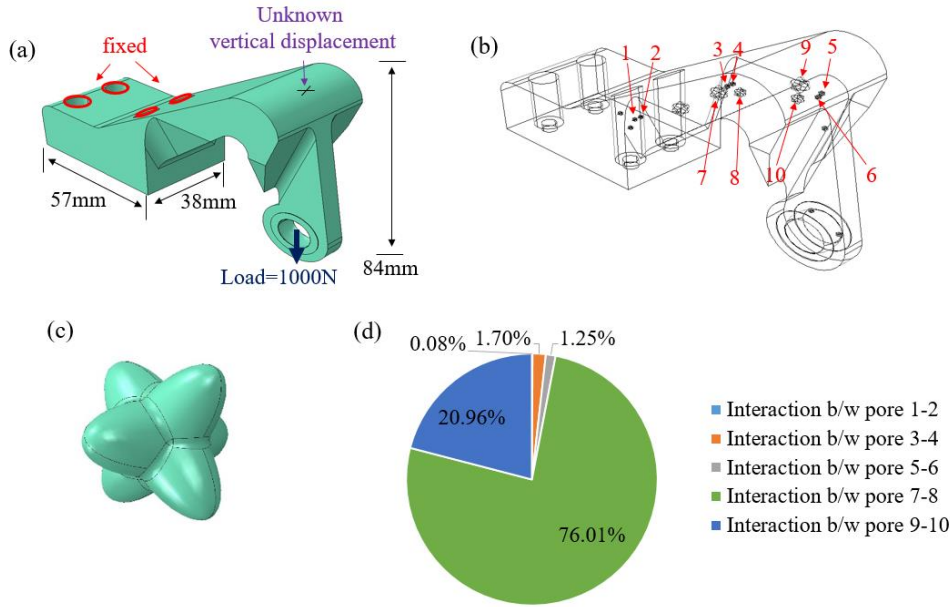

**Figure 13** Cast study on a 3D hook model: (a) Dimensions and boundary conditions, (b) Porosity spatial distributions in the hook model where labeled pores are subject to pairwise interactions, (c) 3D synthetic pore model [13], and (d) Comparison of the pairwise interaction terms among the ten pores.

**Table 4:** Geometries and distances of pores in the 3D hook model.

| Pore index | 1 | 2 | 3 | 4 | 5 | 6 | 7 | 8 | 9 | 10 |
| --- | --- | --- | --- | --- | --- | --- | --- | --- | --- | --- |
| Interactive pore | 2 | 1 | 4 | 3 | 6 | 5 | 8 | 7 | 10 | 9 |
| Diameter (mm) | 2.0 | 2.0 | 2.0 | 2.0 | 2.0 | 2.0 | 6.0 | 4.0 | 6.0 | 4.0 |
| Distance (mm) | 5.94 | 5.94 | 1.73 | 1.73 | 2.24 | 2.24 | 5.92 | 5.92 | 4.69 | 4.69 |

FEA results show the defeaturing error is 3.67E-8 mm due to the presence of pores. The predictions from the first and second-order estimators are 2.47e-8 mm and 3.14e-8 mm, respectively. It is clear the porosity interactions result in higher prediction accuracy from the second-order estimator. We compare the porosity interactions and their relative importance in Figure 13(d), from which several observations can be drawn. First, the interaction between pores 7 and 8 is dominant, accounting for 76.01% of the total interaction. The pair 9-10 with similar pore sizes and distance, comparatively, accounts for only 20.96%; one plausible reason being that the pair 7-8 locates in a highly stressed region. Second, by comparing the pair 3-4 (1.7%) with the pair 7-8, we notice that even though the two pairs locate in the same region, the interaction magnitude in the former pair is much lower due to their smaller sizes. The same observation is noted by comparing the pair 9-10 (20.96%) and 5-6 (1.25%). Lastly, the interaction from pair 1-2 is trivial (0.08%) because of their small sizes and long-distance. It is, therefore, evident from this experiment that for components with many pores, the most important pore-to-pore interactions come from the neighboring pores of large size located in highly stressed regions.

7.4 Case study: 3D tensile bar

The last case study aims to test the performance of the proposed estimator on parts with actual manufacturing pores. A manufactured tensile bar is modeled in this section whose dimensions and loading conditions are described in Figure 14(a). To simulate a tensile test, we fix all degrees of freedom (DOF) on the top and bottom surfaces of gripping sections and apply uniform surface traction as the tensile load. The quantity of interest in this experiment is the pointwise displacement at the center of the cross-section (marked red) along the loading direction.

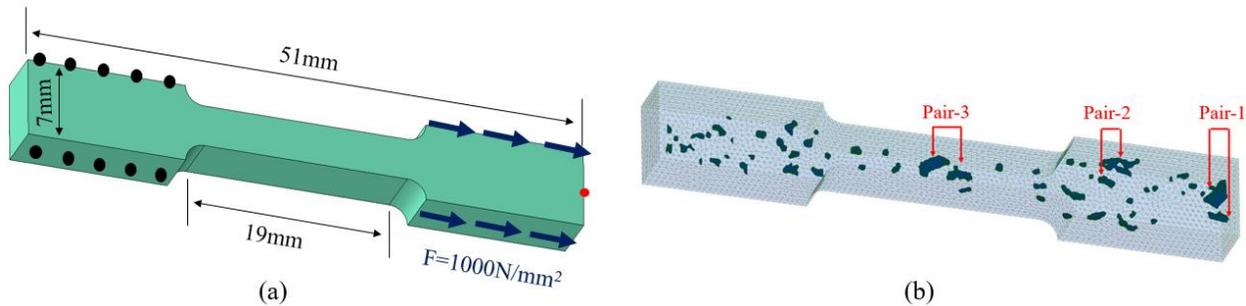

**Figure 14** Case study on a 3D tensile bar: (a) Dimensions and boundary conditions, and (b) Surface representation of the tensile bar with tomography reconstructed porosity. Three pairs of pores are highlighted for the interaction study.

This tensile bar is scanned by an X-ray scanner, VALUCT [43] with a 250 µm resolution. Porosity detection is performed based on the method of thresholding and segmentation between phase contrast in the image processing toolbox in MATLAB [44]. The reconstructed pore geometry is then repaired by SALOME [45] to remove mesh defects, e.g., disconnected geometry, misoriented face, and intersected elements. After mesh clean-up, a surface mesh representation of the porous tensile bar is shown in Figure 14(b). Tomography detects 58 internal pores whose equivalent radii range from 0.54 mm to 3.43mm. Two types of casting pores are observed: gas pores, resulting from trapped gas or released hydrogen from solidified metal, tend to be spherical, and shrinkage pores, due to insufficient fill during solidification, is generally large and irregularly shaped. We note porosity locations generally indicate where the liquid metal is last to be solidified, most pores in this tensile bar distribute on the middle plane along the thickness direction. Among the 58 pores, six pores (i.e., three pairs) are selected for our interaction studies. Such selection is due to their large sizes and close neighbors. Recall from previous experiments that pore sizes and neighbor distance are two major factors in determining interaction magnitudes, and the two factors are summarized in the Table 5.

A volume mesh of 651,814 linear tetrahedral elements is generated from the surface mesh. The volume mesh representation of the porous tensile bar is then solved by a direct FE whose result serves as ground truth to compare with the proposed estimator. Our estimators, on the contrary, only need to be computed on the reference model (no pores) in Figure 14(a). Since meshing the reference model only requires 87,747 linear tetrahedral elements, its mesh size is only about 13.5% of the direct FEA approach. The defeaturing error from FEA shows a displacement of 3.54e-8 mm with the presence of pores. The predictions from the first and second-order estimators are 1.63e-8 mm and 2.79e-8 mm, respectively. It is thus evident the higher-order estimator is more accurate in this case. The values of the proposed estimators are detailed for each pair in Table 5. It should be noted the sum of $\mathcal{D}^1_{pore}$ does not equal the value of the first-order estimation because this estimation also includes defeaturing errors on other pores whose interactions are not included; the same applies to $\mathcal{D}^2_{pore}$. From this table, we can see pair-3 has higher estimated interaction than others, although the three pairs have similar sizes and distances. One plausible reason is that pair-3 locates in the gauge section with the reduced cross-sectional area and high stress concentrations (see Equation (6.1) and (6.2)). The percentages of interactions in estimators range from 1.95% to 3.62%. Stronger interactions would be expected if pore distances are even smaller.

**Table 5:** Pore geometries and estimator values for the 3D cantilever beam model in Figure 14(b).

|  | Average size (mm) | Distance (mm) | $\mathcal{D}^1_{pore}$ | $\mathcal{D}^2_{pore}$ | $\mathcal{D}^2_{int}$ | $\mathcal{D}^2_{int}/\mathcal{D}^2_{pore}$ |
|---|---|---|---|---|---|---|
| Pair-1 | 3.13 | 2.47 | 4.67E-12 | 9.25E-12 | 3.35E-13 | 3.62% |
| Pair-2 | 2.97 | 1.75 | 8.01E-11 | 1.52E-10 | 2.42E-12 | 1.95% |
| Pair-3 | 3.18 | 2.89 | 7.04E-9 | 8.92E-9 | 2.01E-10 | 2.25% |

Computational efficiency is illustrated in this example by comparing against FEM from two perspectives. First, in a pre-processing stage, mesh generation of our method is much more efficient than FEM which generally requires high-quality discretization. Although FEM would be dramatically accelerated via meshing automation, the automation becomes difficult for geometries with complex features. The cast tensile bar, for example, contains manufacturing-induced pores of complex morphologies. Without human intervention, meshing singularity is not rare. Problematic mesh results in ill-conditioned stiffness matrices, deteriorating computational accuracy and efficiency. Our method, comparatively, allows performing on the reference domain (simple geometry and no pores) with efficient mesh generation. Second, in solving process, the number of unknown variables in our method is only 13.5% of the direct FEM, leading to much smaller memory footprints and faster execution. With the two perspectives, we compare the computational time between the direct FEM and our method in Table 6. Our method reduces the computational time by 82%. Specifically, in the meshing step, most FEM time comes from human interventions to fix mesh singularities, e.g., misoriented faces, missing nodes, and ill-shaped volume elements due to the sharp corners close to porosity. In the solving process, solving direct FEM takes 243 seconds on the porous domain discretized by 651,814 elements. Our method, in comparison, spends 55 seconds in solution on the dense domain meshed with 87,747 elements and 46 seconds in solving exterior Neumann formulations for the first and second-order estimators.

**Table 6:** Time comparison (in seconds) on different steps between direct FEM and the proposed estimator.

|  | Direct FEM | Proposed |
| --- | --- | --- |
| Pore reconstruction | 19 | 19 |
| Meshing | 1210 | 2 |
| Solving | 243 | 101 |
| Total time (sec) | 1472 | 265 |

In this section, we demonstrate our method via several numerical experiments. It is first tested on benchmark examples where the influences of various pore parameters are studied. It is then tested on different cast studies. In the bracket and hook models, porosity spatial distributions are predicted via casting simulations, and pore shapes are represented by synthetic models. In the tensile bar model, pores are induced by casting processes and their locations and morphologies are reconstructed via computed tomography. Through the benchmarks and applications, we have observed: (1) The proposed gradient-enhanced estimator, as a reliable and computationally efficient alternative to direct FEM, is generally more accurate than the first-order estimator and the topological sensitivity-based estimator in conditions with strong inter-pore interactions; (2) Pore-to-pore interactions are dependent on the ratio between inter-pore distances and pore sizes, and other important factors include pore shapes, locations, and local field variables; (3) For a metallic component containing multiple pores, interactions between closely distanced large pores in highly stressed regions are more critical than others, which needs careful attention. Although the approach is tested and verified by casting pores in numerical experiments, its methodology is developed without a limitation to a specific manufacturing type. Its extension to other process-induced porosity problems (e.g., additive manufacturing) is, therefore, possible but needs to be established.

## 8. CONCLUSIONS

In this paper, a novel second-order defeaturing porosity estimator is proposed to predict the elastic structural performance with respect to manufacturing-induced pores. The proposed estimator, which combines the merits from topological sensitivity, the first and second-order shape sensitivities, is advantageous in accounting for complex porosity morphologies and pore-to-pore interactions. Its advantage over classic FEM is that it only requires solutions from the reference domains without pores, thus preventing mesh singularities and lowering computational costs.

There are several directions to expand the current work in the future. First, this method is currently only applied to predict the quantity of interests for elastic problems. It is well known both shape and topological sensitivities are well developed for nonlinear problems [23,24]. Extension to nonlinear problems is of great interest. For instance, fatigue analyses on porosity materials are arguably more important [11]. Second, material properties in this work are assumed to be isotropic. This assumption can be relaxed in future work to consider more realistic engineered alloys with heterogeneous properties. Third, compared to pores in casting alloys, the pores in additive manufactured alloys may have very different morphology and spatial distribution due to different formation mechanisms [13]. Applying the proposed method to study additive manufacturing-induced porosity has much value and its prediction accuracy needs to be investigated.

**ACKNOWLEDGEMENTS**

The authors thank the ACRC consortium members for funding the project and also thank the industrial members of the focus group whose help was invaluable. Specifically, we appreciate Randy Beals from Magna International provided the W-profile plate samples for testing, and Chen Dai from VJ Technologies provided support on X-ray computed tomography (CT) scanning and data generation.

**Appendix A. Topology Sensitivity**

The proposed porosity estimator is an extension to topological sensitivity whose concept is reviewed. Topological sensitivity captures the first-order impact of inserting an infinitesimally small spherical hole within a domain on various quantities of interests [32–39].

Let a quantity of interest $\Psi_0(\Omega_s)$ be defined within a region of interests $\Omega_s$ on a smooth bounded domain $\Omega$ in Figure A.1. Suppose an infinitesimally small hole with the radius ($\xi$) is introduced by perturbing the domain $\Omega$ at an arbitrary location $\hat{\mathbf{x}}$. A new domain $(\Omega - \Omega_p^\xi)$ is generated with the boundary $(\Gamma + \Gamma_p^\xi)$. Given the topological change, the performance function $\Psi_\xi(\Omega_s)$ defined in the same region $\Omega_s$ but on the perturbed domain can be written as:

$$\Psi_\xi(\Omega_S) = \Psi_0(\Omega_S) + g(\xi)\mathcal{T}_{topo}(\hat{\mathbf{x}}) + R(g(\xi)) \tag{A.1}$$

where $g(\xi)$ is a monotone function, which decreases to zero as the size of the pore ($\xi$) approaches zero. $T_{topo}$ is the first-order topological derivative defined at the pore, and $R$ contains all higher-order terms. Drop the higher-order terms ($R$) in Equation (A.1), and rewrite it as the classic topological sensitivity as:

$$\mathcal{T}_{topo}(\hat{\mathbf{x}}) \equiv \lim_{\xi \to 0} \frac{\Psi_\xi(\Omega_S) - \Psi_0(\Omega_S)}{g(\xi)} \tag{A.2}$$

where the monotone function $g(\xi)$ is taken as the volume of the small spherical pore [33]. To find the closed-form expression for the topological sensitivity, we must first define an adjoint. Recall that the adjoint field associated with the quantity of interest ($\Psi$) satisfies [23,40]:

$$\mathbf{K}\boldsymbol{\lambda} = -\nabla_\mathbf{z}\Psi \tag{A.3}$$

where $\mathbf{K}$ is the linear elastic stiffness matrix, $\mathbf{z}$ is the primary displacement solution, and $\boldsymbol{\lambda}$ is the adjoint solution. Its right side represents the adjoint loads, which can be determined via distribution theory [54]. After the adjoint solution $\boldsymbol{\lambda}$ is calculated, the topological derivative [33] can be computed at an arbitrary point over the reference domain (without pores) as:

$$\mathcal{T}_{topo}(\hat{\mathbf{x}}) = \frac{3}{4}\frac{1-\nu}{7-5\nu}\left[10\boldsymbol{\sigma}(\mathbf{z}):\boldsymbol{\varepsilon}(\boldsymbol{\lambda}) - \frac{1-5\nu}{1-2\nu}tr[\boldsymbol{\sigma}(\mathbf{z})]tr[\boldsymbol{\varepsilon}(\boldsymbol{\lambda})]\right] \tag{A.4}$$

where $\nu$ is the Poisson ratio, $\boldsymbol{\sigma}(\mathbf{z})$ and $\boldsymbol{\varepsilon}(\boldsymbol{\lambda})$ are the stress and strain computed from the primary ($\mathbf{z}$) and adjoint solutions ($\boldsymbol{\lambda}$). The performance function ($\Psi$) on the porous domain with an infinitesimal spherical pore can be estimated by multiplying the sensitivity with the volume of the pore as:

$$\Psi(\Omega_S) = \Psi_0(\Omega_S) + Vol(\xi)\mathcal{T}_{topo}(\hat{\mathbf{x}}) \tag{A.5}$$

where $Vol(\xi)$ is the pore's volume. Therefore, the topology sensitivity-based porosity estimator is:

$$\mathcal{D}_{topo} = \Psi(\Omega_S) - \Psi_0(\Omega_S) = Vol(\xi)\mathcal{T}_{topo}(\hat{\mathbf{x}}) \tag{A.6}$$

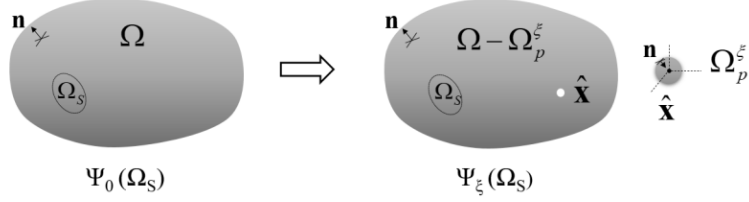

**Figure A.1:** Illustration of a topological change in a smooth 3D domain: the reference domain ($\Omega$) is topologically perturbed by introducing an infinitesimal spherical pore ($\Omega_p^\xi$) at $\hat{\mathbf{x}}$ with a radius of $\xi$. The first-order change of the performance function $\Psi$ over the fixed region $\Omega_s$ is captured by the topology sensitivity.

## Appendix B. First-Order Shape Sensitivity

For a linear elastic material with constitutive defined in Equation (B.1), its strain $\boldsymbol{\varepsilon}$, strain energy $U(\mathbf{z})$, external load potential $E(\mathbf{z})$ and, total potential energy $\Pi(\mathbf{z})$ are defined in Equations (B.2)-(B.5).

$$\boldsymbol{\sigma} = \mathbb{C}:\boldsymbol{\varepsilon} \tag{B.1}$$

$$\boldsymbol{\varepsilon} = \frac{1}{2}(\nabla\mathbf{z} + \nabla\mathbf{z}^{\mathbf{T}}) \tag{B.2}$$

$$U(\mathbf{z}) = \frac{1}{2}\iiint_\Omega \boldsymbol{\sigma}(\mathbf{z}):\boldsymbol{\varepsilon}(\mathbf{z})d\Omega \tag{B.3}$$

$$E(\mathbf{z}) = \iiint_\Omega \mathbf{z}^{\mathbf{T}}\mathbf{f}^{\mathbf{b}}d\Omega + \iint_{\Gamma^s}\mathbf{z}^{\mathbf{T}}\mathbf{f}^{\mathbf{s}}d\Gamma \tag{B.4}$$

$$\Pi(\mathbf{z}) = U(\mathbf{z}) - E(\mathbf{z}) \tag{B.5}$$

where $\mathbb{C}$ is the elastic tangent matrix, $\boldsymbol{\sigma}$ is stress, $\mathbf{z}$ is the primary (displacement) solution, $\boldsymbol{f}^b$ and $\boldsymbol{f}^S$ are body force and surface traction, respectively. The primary solution ($\mathbf{z}$) is computed by minimizing the total potential energy in a variational form as:

$$\delta\Pi(\mathbf{z},\bar{\mathbf{z}}) = 0 = \delta U(\mathbf{z},\bar{\mathbf{z}}) - \delta E(\mathbf{z},\bar{\mathbf{z}}) \tag{B.6}$$

in which

$$a(\mathbf{z},\bar{\mathbf{z}}) \equiv \delta U(\mathbf{z},\bar{\mathbf{z}}) = \iiint_\Omega \boldsymbol{\sigma}(\mathbf{z}):\boldsymbol{\varepsilon}(\bar{\mathbf{z}})d\Omega \tag{B.7}$$

$$l(\bar{\mathbf{z}}) \equiv \delta E(\mathbf{z},\bar{\mathbf{z}}) = \iiint_\Omega \bar{\mathbf{z}}^{\mathbf{T}}\mathbf{f}^{\mathbf{b}}d\Omega + \iint_{\Gamma^s}\bar{\mathbf{z}}^{\mathbf{T}}\mathbf{f}^{\mathbf{s}}d\Gamma \tag{B.8}$$

where $a(\mathbf{z},\bar{\mathbf{z}})$ and $l(\bar{\mathbf{z}})$ are the variational forms of strain energy and external load potential, respectively, and $\bar{\mathbf{z}}$ is an arbitrary field in a Hilbert space satisfying the kinematically admissible condition in Equation (B.9). The primary solution ($\mathbf{z}$) is then computed by finding a solution ($\mathbf{z} \in Z$) to Equation (B.10).

$$Z = \{\mathbf{z} \in H^1(\Omega) \mid \mathbf{z} = \mathbf{0} \text{ on } \Gamma^h\} \tag{B.9}$$

$$a(\mathbf{z},\bar{\mathbf{z}}) = l(\bar{\mathbf{z}}) \tag{B.10}$$

By taking material derivative (see Section 9) of Equation (B.10) and assuming external forces are independent of shape changes, we compute the derivatives of the variational equations as:

$$a'(\mathbf{z}, \bar{\mathbf{z}}) = l'(\bar{\mathbf{z}}) \tag{B.11}$$

$$a'(\mathbf{z}, \bar{\mathbf{z}}) = \iiint_\Omega [\boldsymbol{\varepsilon}(\bar{\mathbf{z}}'):\boldsymbol{\sigma}(\mathbf{z}) + \boldsymbol{\varepsilon}(\bar{\mathbf{z}}):\boldsymbol{\sigma}(\mathbf{z}')]d\Omega + \iint_\Gamma [\boldsymbol{\varepsilon}(\bar{\mathbf{z}}):\boldsymbol{\sigma}(\mathbf{z})]V_n d\Gamma \tag{B.12}$$

$$l'(\bar{\mathbf{z}}) = \iiint_\Omega \mathbf{f}^{\mathbf{bT}}\bar{\mathbf{z}}'d\Omega + \iint_{\Gamma^{f+s}} (\mathbf{f}^{\mathbf{bT}}\bar{\mathbf{z}})V_n d\Gamma + \iint_{\Gamma^s} \left\{ \mathbf{f}^{\mathbf{sT}}\bar{\mathbf{z}}' + \left[\nabla(\mathbf{f}^{\mathbf{sT}}\bar{\mathbf{z}})^T \mathbf{n} + \Lambda(\mathbf{f}^{\mathbf{sT}}\bar{\mathbf{z}})\right]\right\}V_n d\Gamma \tag{B.13}$$

where both $\dot{\mathbf{z}}$ and $\bar{\mathbf{z}}$ satisfy the kinematic admissible boundary condition in Equation (B.9). To compute porosity shape sensitivity, we define a generic elastic performance function ($\Psi$) on the porous domain in Figure 4(a) as:

$$\Psi = \iiint_{\Omega_S} g(\mathbf{x})d\Omega \tag{B.14}$$

where g(**x**) is a scalar function defined over a fixed region of interest $\Omega_s$ with material points **x**. For example, a pointwise displacement at a point $\hat{\mathbf{x}}$ can be defined as the performance function in Equation (B.15) with its material derivative in Equation (B.16). The adjoint solution ($\boldsymbol{\lambda}$) is defined in Equation (B.17).

$$\Psi = \iiint_\Omega \delta(\mathbf{x}-\hat{\mathbf{x}})\mathbf{z}d\Omega \tag{B.15}$$

$$\Psi' = \iiint_\Omega \delta(\mathbf{x}-\hat{\mathbf{x}})(\dot{\mathbf{z}} - \nabla\mathbf{z}^T\mathbf{V})d\Omega \tag{B.16}$$

$$a(\boldsymbol{\lambda}, \bar{\boldsymbol{\lambda}}) = \iiint_\Omega \delta(\mathbf{x}-\hat{\mathbf{x}})\bar{\boldsymbol{\lambda}}d\Omega \tag{B.17}$$

where **V** is the design speed defined in Equation (3.4), and $\bar{\boldsymbol{\lambda}}$ is an arbitrary adjoint field in the Hilbert space. By replacing the virtual displacement $\bar{\boldsymbol{\lambda}}$ by $\dot{\mathbf{z}}$ in Equation (B.17) and combining with Equations (B.11) and (B.16), we obtain the shape sensitivity of the pointwise displacement in volume integral form in Equation (B.18).

$$\begin{aligned}\Psi' = &\iiint_\Omega \begin{bmatrix} \boldsymbol{\sigma}(\mathbf{z}):\boldsymbol{\varepsilon}(\nabla\boldsymbol{\lambda}^T\mathbf{V}) + \boldsymbol{\sigma}(\nabla\mathbf{z}^T\mathbf{V}):\boldsymbol{\varepsilon}(\boldsymbol{\lambda}) \\ -\delta(\mathbf{x}-\hat{\mathbf{x}})^T(\nabla\boldsymbol{\lambda}^T\mathbf{V}) - \delta(\mathbf{x}-\hat{\mathbf{x}})^T(\nabla\mathbf{z}^T\mathbf{V}) \end{bmatrix} d\Omega \\ &- \iint_\Gamma [\boldsymbol{\sigma}(\mathbf{z}):\boldsymbol{\varepsilon}(\boldsymbol{\lambda})]V_n d\Gamma + \iint_{\Gamma^{f+s}} \left[\delta(\mathbf{x}-\hat{\mathbf{x}})^T \boldsymbol{\lambda}\right]V_n d\Gamma \\ &+ \iint_{\Gamma^s} \left\{-\mathbf{f}^{\mathbf{sT}}(\nabla\boldsymbol{\lambda}^T\mathbf{V}) + \left[\nabla(\mathbf{f}^{\mathbf{sT}}\boldsymbol{\lambda})\cdot\mathbf{n} + \Lambda(\mathbf{f}^{\mathbf{sT}}\boldsymbol{\lambda})\right]V_n\right\}d\Gamma \end{aligned} \tag{B.18}$$

where $V_n$ is the normal design speed defined over pore surfaces pointing in normal directions (**n**), $\Gamma^s$ is the surface prescribed with tractions, and $\Lambda$ is the mean of surface curvatures. It is noteworthy that the sensitivity's volume form is well suited for a general-purpose design sensitivity analysis, e.g., FEM. Its boundary form, however, is preferred in this work for the following reasons: (1) tomography reconstruction provides a detailed geometric description of pore surfaces, which together with domain perturbations gives an excellent explanation of design velocity, and (2) generation of 2D high-fidelity mesh is much more efficient and easier than a 3D volume mesh. To obtain the boundary form, we rely on the variational identities of the primary and adjoint solutions in Equations (B.19) and (B.20).

$$\iiint_\Omega \boldsymbol{\sigma}(\mathbf{z}):\boldsymbol{\varepsilon}(\bar{\mathbf{z}})d\Omega - \iiint_\Omega \mathbf{f}^{\mathbf{bT}}\bar{\mathbf{z}}d\Omega = \iint_\Gamma \bar{\mathbf{z}}^T(\boldsymbol{\sigma}(\mathbf{z})\cdot\mathbf{n})d\Gamma \tag{B.19}$$

$$\iiint_\Omega \boldsymbol{\sigma}(\lambda):\boldsymbol{\varepsilon}(\bar{\lambda})d\Omega - \iiint_\Omega \bar{\lambda}^T \delta(\mathbf{x}-\hat{\mathbf{x}})d\Omega = \iint_\Gamma \bar{\lambda}^T (\boldsymbol{\sigma}(\lambda)\cdot\mathbf{n})d\Gamma \tag{B.20}$$

By plugging the two variational identities into the domain form in Equation (B.18) and dropping all terms without design speeds, the porosity shape sensitivity can be expressed in its boundary form as:

$$\Psi' = \frac{d(\Psi(\Omega_s))}{d\eta} = -\iint_{\Gamma^p} [\boldsymbol{\sigma}(\mathbf{z}):\boldsymbol{\varepsilon}(\lambda)]V_n d\Gamma \tag{B.21}$$

where $\Gamma^p$ refers to pore surfaces with normal design speeds $V_n$, $\boldsymbol{\sigma}(\mathbf{z})$ and $\boldsymbol{\varepsilon}(\lambda)$ are the stress and strain on pore surface computed by primary and adjoint solutions, and η is the pore shape parameter (see Section 3). We point out since design speed is only defined over the boundary of pores, the boundary integration therefore only needs to be performed on pore surfaces.

**Appendix C. Second-Order Shape Sensitivity**

To compute the second-order porosity sensitivity on a domain with two closely-distanced pores, we start with the second-order Lagrangian as:

$$\begin{aligned}&L_2(t,s;\mathbf{z},\lambda,\mathbf{P},\mathbf{Q}) \\ &= \frac{\partial L(t,s;\mathbf{z},\lambda)}{\partial t} + [a(t,s;\mathbf{z},\mathbf{P}) - l(t,s;\mathbf{P})] + [\langle D_3 a(t,s;\mathbf{z},\lambda);\mathbf{Q}\rangle + \langle D_3 \Psi(t,s;\mathbf{z});\mathbf{Q}\rangle]\end{aligned} \tag{C.1}$$

where $L_2$ is the Lagrangian function, $t$ and $s$ are two pore shape parameters, $\mathbf{z}$ and $\lambda$ are primary and adjoint solutions, $\Psi$ is the performance function (e.g., the pointwise displacement in Equation (B.15)), $\mathbf{P}$ and $\mathbf{Q}$ are the two Lagrangian multipliers, $\langle D\rangle$ is the Gateaux derivative (see Section 4), and $a$ and $l$ are the variational forms of strain energy and external load potential, respectively. We note that the second term on the right side presents the constraint of Equation (4.8) and the third term is associated with the constraint of adjoints in Equation (B.17). Computing the second-order sensitivity boils down to taking derivative of the second-order Lagrangian in an open interval $s \in (-\tau_s, +\tau_s)$ as [34]:

$$\frac{\partial^2 \Psi}{\partial t \partial s} = \frac{\partial L_2(t,s;\mathbf{z},\lambda,\mathbf{P},\mathbf{Q})}{\partial s} \tag{C.2}$$

To compute the two multipliers ($\mathbf{P}$ and $\mathbf{Q}$), we start to evaluate $\mathbf{Q}$ by solving:

$$\langle D_3 a(t,s;\mathbf{z},\lambda);\mathbf{Q}\rangle = \frac{\partial l}{\partial t}(t,s;\lambda) - \frac{\partial a}{\partial t}(t,s;\mathbf{z},\lambda) \tag{C.3}$$

where $\langle D\rangle$ is the Gateaus derivative in Equation (4.11). To compute the sensitivity to the pore parameters ($t$ and $s$) with design speeds $\mathbf{V}$ and $\mathbf{W}$, we transform all terms in Equation (C.3) to the parameterized domain ($\Omega_\eta$) by setting ($t=s=0$) as:

$$\langle D_3 a(t,s;\mathbf{z},\lambda);\mathbf{Q}\rangle|_{t=s=0} = \frac{1}{4}\iiint_{\Omega_\eta}[(\nabla\lambda + \nabla\lambda^T):\mathbb{C}:(\nabla\mathbf{Q}+\nabla\mathbf{Q}^T)]d\Omega_\eta = \iiint_{\Omega_\eta} \boldsymbol{\varepsilon}(\lambda):\boldsymbol{\sigma}(\mathbf{Q})d\Omega_\eta \tag{C.4}$$

$$\frac{\partial l}{\partial t}(t,s;\lambda)\bigg|_{t=s=0} = \iiint_{\Omega_\eta}(\lambda^T \mathbf{f}^b)\nabla\cdot\mathbf{V}d\Omega_\eta \tag{C.5}$$

$$\left.\frac{\partial a}{\partial t}(t,s;\mathbf{z},\boldsymbol{\lambda})\right|_{t=s=0} = \iiint_{\Omega_\eta} \left[\begin{array}{l}\boldsymbol{\varepsilon}(\boldsymbol{\lambda}):\boldsymbol{\sigma}(\mathbf{z})(\nabla\cdot\mathbf{V})-\frac{1}{2}\left(\nabla\mathbf{V}^T\nabla\boldsymbol{\lambda}+\nabla\boldsymbol{\lambda}^T\nabla\mathbf{V}\right):\boldsymbol{\sigma}(\mathbf{z}) \\ -\frac{1}{2}\boldsymbol{\sigma}(\boldsymbol{\lambda}):\left(\nabla\mathbf{V}^T\nabla\mathbf{z}+\nabla\mathbf{z}^T\nabla\mathbf{V}\right)\end{array}\right]d\Omega_\eta \quad (\text{C.6})$$

By combining Equations (C.4), (C.5) and (C.6) with Equation (C.3), we obtain Equation (C.7) which is further simplified by the theory of distribution to Equation (C.8).

$$\iiint_{\Omega_\eta} \boldsymbol{\varepsilon}(\boldsymbol{\lambda}):\boldsymbol{\sigma}(\mathbf{Q})d\Omega_\eta = \iint_{\Gamma_\eta} \left(\boldsymbol{\sigma}(\boldsymbol{\lambda})\nabla\mathbf{z}^T\mathbf{V}\right)\cdot\mathbf{n}d\Gamma_\eta \quad (\text{C.7})$$

$$\iiint_{\Omega_\eta} \boldsymbol{\varepsilon}(\boldsymbol{\lambda}):\boldsymbol{\sigma}(\mathbf{Q})d\Omega_\eta = -\iint_{\Gamma_\eta} \boldsymbol{\lambda}\cdot\left[\nabla\cdot\mathbb{C}:\left(\frac{1}{2}\left(\mathbf{A}+\mathbf{A}^T\right)\right)\right]d\Gamma_\eta, \quad \mathbf{A}=\mathbf{n}\otimes\left(\nabla\mathbf{z}^T\mathbf{V}\right) \quad (\text{C.8})$$

Evaluation of $\mathbf{P}$ requires to expand Equation (C.9) to Equations (C.10)-(C.12) as:

$$\langle D_3 a(t,s;\mathbf{z},\mathbf{P});\mathbf{v}\rangle = -\langle D_{13}^2\Psi(t,s;\mathbf{z});\mathbf{v}\rangle\Big|_{t=s=0} - \langle D_{13}^2 a(t,s;\mathbf{z},\boldsymbol{\lambda});\mathbf{v}\rangle\Big|_{t=s=0}$$
$$-\langle D_{33}^2 a(t,s;\mathbf{z},\boldsymbol{\lambda}):(\mathbf{Q},\mathbf{v})\rangle\Big|_{t=s=0} - \langle D_{33}^2\Psi(t,s;\mathbf{z}):(\mathbf{Q},\mathbf{v})\rangle\Big|_{t=s=0} \quad (\text{C.9})$$

$$\langle D_3 a(t,s;\mathbf{z},\mathbf{P});\mathbf{v}\rangle = \iiint_{\Omega_\eta} \boldsymbol{\varepsilon}(\mathbf{P}):\boldsymbol{\sigma}(\mathbf{v})d\Omega_\eta \quad (\text{C.10})$$

$$\langle D_{13}^2\Psi(t,s;\mathbf{z});\mathbf{v}\rangle\Big|_{t=s=0} = \iiint_{\Omega_{0\eta}} \delta(\mathbf{x}-\hat{\mathbf{x}})^T\mathbf{v}(\nabla\cdot\mathbf{V})d\Omega_\eta \quad (\text{C.11})$$

$$\langle D_{13}^2 a(t,s;\mathbf{z},\boldsymbol{\lambda});\mathbf{v}\rangle\Big|_{t=s=0} = \iiint_{\Omega_\eta} \left\{\begin{array}{l}(\boldsymbol{\varepsilon}(\boldsymbol{\lambda}):\boldsymbol{\sigma}(\mathbf{v}))\nabla\cdot\mathbf{V}-\frac{1}{2}\left(\nabla\mathbf{V}^T\nabla\boldsymbol{\lambda}+\nabla\boldsymbol{\lambda}^T\nabla\mathbf{V}\right):\boldsymbol{\sigma}(\mathbf{v}) \\ -\frac{1}{2}\boldsymbol{\sigma}(\boldsymbol{\lambda}):\left(\nabla\mathbf{V}^T\nabla\mathbf{v}+\nabla\mathbf{v}^T\nabla\mathbf{V}\right)\end{array}\right\}d\Omega_\eta \quad (\text{C.12})$$

where $\mathbf{v}$ is an arbitrary field on the parameterized domain ($\mathbf{v}\in\mathbf{U}(\Omega_\eta)$). We note that since both $a(t,s;\mathbf{z},\boldsymbol{\eta})$ and $\Psi(t,s;\mathbf{z})$ are linear of the variable $\mathbf{z}$, their second-order derivatives vanish:

$$\langle D_{33}^2 a(t,s;\mathbf{z},\boldsymbol{\lambda}):(\mathbf{Q},\mathbf{v})\rangle\Big|_{t=s=0} = 0 \quad (\text{C.13})$$

$$\langle D_{33}^2\Psi(t,s;\mathbf{z}):(\mathbf{Q},\mathbf{v})\rangle\Big|_{t=s=0} = 0 \quad (\text{C.14})$$

By combining Equations (C.10)-(C.14) with Equation (C.9), we evaluate the multiplier $\mathbf{P}$ by:

$$\iiint_{\Omega_\eta} \boldsymbol{\varepsilon}(\mathbf{P}):\boldsymbol{\sigma}(\mathbf{v})d\Omega_\eta = -\iint_{\Gamma_\eta} \mathbf{v}\cdot\left[\nabla\cdot\mathbb{C}:\left(\frac{1}{2}\left(\mathbf{B}+\mathbf{B}^T\right)\right)\right]d\Gamma_\eta, \quad \mathbf{B}=\mathbf{n}\otimes\left(\nabla\boldsymbol{\lambda}^T\mathbf{V}\right) \quad (\text{C.15})$$

To obtain the second-order shape sensitivity, the domain integral of $<D<D\Psi;\boldsymbol{V}>;\boldsymbol{W}>$ needs to be computed in Equation (C.16). Its first term on the right side can be expanded in Equation (C.17).

$$\langle D\langle D\psi;\mathbf{V}\rangle;\mathbf{W}\rangle = \langle D^2\psi:(\mathbf{V},\mathbf{W})\rangle + \langle D\psi;(\nabla\mathbf{V})\mathbf{W}\rangle \quad (\text{C.16})$$

$$\langle D^2\psi:(\mathbf{V},\mathbf{W})\rangle = \left.\frac{\partial^2\Psi}{\partial t\partial s}(t,s;\mathbf{z})\right|_{t=s=0} + \left.\frac{\partial^2 a}{\partial t\partial s}(t,s;\mathbf{z},\boldsymbol{\lambda})\right|_{t=s=0} - \left.\frac{\partial^2 l}{\partial t\partial s}(t,s;\boldsymbol{\lambda})\right|_{t=s=0}$$
$$+\left.\frac{\partial a}{\partial s}(t,s;\mathbf{z},\mathbf{P})\right|_{t=s=0} - \left.\frac{\partial l}{\partial s}(t,s;\mathbf{P})\right|_{t=s=0} + \langle D_{23}^2 a(t,s;\mathbf{z},\boldsymbol{\lambda});\mathbf{Q}\rangle\Big|_{t=s=0} + \langle D_{23}^2\Psi(t,s;\mathbf{z});\mathbf{Q}\rangle\Big|_{t=s=0} \quad (\text{C.17})$$

Each term of Equation (C.17) is further expanded in the forms of domain integral as:

$$\left.\frac{\partial \Psi}{\partial t}(t,s;\mathbf{z})\right|_{t=s=0} = \iiint_{\Omega_\eta} \left[\delta(\mathbf{x}-\hat{\mathbf{x}})\mathbf{z}\right] \nabla \cdot \mathbf{V} d\Omega_\eta \qquad (C.18)$$

$$\left.\frac{\partial^2 \Psi}{\partial t \partial s}(t,s;\mathbf{z})\right|_{t=s=0} = \mathbf{0} \qquad (C.19)$$

$$\left.\frac{\partial l}{\partial t}(t,s;\boldsymbol{\lambda})\right|_{t=s=0} = \iiint_{\Omega_\eta} \left(\boldsymbol{\lambda}^T \mathbf{f}^b\right) \nabla \cdot \mathbf{V} d\Omega_\eta \qquad (C.20)$$

$$\left.\frac{\partial^2 l}{\partial t \partial s}(t,s;l)\right|_{t=s=0} = \mathbf{0} \qquad (C.21)$$

$$\left.\frac{\partial a}{\partial s}(t,s;\mathbf{z},\mathbf{P})\right|_{t=s=0} = \iiint_{\Omega_\eta} \left(\boldsymbol{\varepsilon}(\mathbf{P}):\boldsymbol{\sigma}(\mathbf{z})\right) \nabla \cdot \mathbf{W} d\Omega_\eta - \frac{1}{2}\iiint_{\Omega_\eta} \left(\nabla \mathbf{W}^T \nabla \mathbf{P} + \nabla \mathbf{P}^T \nabla \mathbf{W}\right) : \boldsymbol{\sigma}(\mathbf{z}) d\Omega_\eta$$
$$-\frac{1}{2}\iiint_{\Omega_\eta} \boldsymbol{\sigma}(\mathbf{P}): \left(\nabla \mathbf{W}^T \nabla \mathbf{z} + \nabla \mathbf{z}^T \nabla \mathbf{W}\right) d\Omega_\eta \qquad (C.22)$$

$$\left.\frac{\partial l}{\partial s}(t,s;\mathbf{P})\right|_{t=s=0} = \iiint_{\Omega_\eta} \left(\mathbf{P}^T \mathbf{f}^b\right) \nabla \cdot \mathbf{W} d\Omega_\eta \qquad (C.23)$$

$$\left.\left\langle D_{23}^2 a(t,s;\mathbf{z},\boldsymbol{\lambda});\mathbf{Q}\right\rangle\right|_{t=s=0} = \iiint_{\Omega_\eta} \begin{bmatrix} \left(\boldsymbol{\varepsilon}(\boldsymbol{\lambda}):\boldsymbol{\sigma}(\mathbf{Q})\right) \nabla \cdot \mathbf{W} - \frac{1}{2}\left(\nabla \mathbf{W}^T \nabla \boldsymbol{\lambda} + \nabla \boldsymbol{\lambda}^T \nabla \mathbf{W}\right):\boldsymbol{\sigma}(\mathbf{Q}) \\ -\frac{1}{2}\boldsymbol{\sigma}(\boldsymbol{\lambda}):\left(\nabla \mathbf{W}^T \nabla \mathbf{Q} + \nabla \mathbf{Q}^T \nabla \mathbf{W}\right) \end{bmatrix} d\Omega_\eta \qquad (C.24)$$

$$\left.\left\langle D_{23}^2 \Psi(t,s;\mathbf{z});\mathbf{Q}\right\rangle\right|_{t=s=0} = \iiint_{\Omega_\eta} \left[\delta(\mathbf{x}-\hat{\mathbf{x}})^T \mathbf{Q}\right] \nabla \cdot \mathbf{W} d\Omega_\eta \qquad (C.25)$$

The second term on the right of Equation (C.16) is also written in a volume form as:

$$\langle D\psi;(\nabla \mathbf{V})\mathbf{W}\rangle$$

$$= \iiint_{\Omega_\eta} \left[\delta(\mathbf{x}-\hat{\mathbf{x}})^T \mathbf{z}\right] \nabla \cdot ((\nabla \mathbf{V})\mathbf{W}) d\Omega_\eta + \iiint_{\Omega_\eta} \begin{Bmatrix} \left[\boldsymbol{\varepsilon}(\boldsymbol{\lambda}):\boldsymbol{\sigma}(\mathbf{z}) - \boldsymbol{\lambda}^T \mathbf{f}^b\right]\mathbf{I} \\ -\nabla \boldsymbol{\lambda}:\boldsymbol{\sigma}(\mathbf{z}) - \nabla \mathbf{z}:\boldsymbol{\sigma}(\boldsymbol{\lambda}) \end{Bmatrix} \nabla \cdot ((\nabla \mathbf{V})\mathbf{W}) d\Omega_\eta \qquad (C.26)$$

$$= \iiint_{\Omega_\eta} \left[\delta(\mathbf{x}-\hat{\mathbf{x}})^T \mathbf{z} + \mathrm{N}\right]\Theta d\Omega_\eta + \iiint_{\Omega_\eta} \left[\left(\delta(\mathbf{x}-\hat{\mathbf{x}})^T \mathbf{z}\right)\nabla \mathbf{V} + \mathrm{N}\nabla \mathbf{V}\right] : \nabla \mathbf{W} d\Omega_\eta$$

with

$$\mathrm{N} = \boldsymbol{\varepsilon}(\boldsymbol{\lambda}):\boldsymbol{\sigma}(\mathbf{z}) - \boldsymbol{\lambda}^T \mathbf{f}^b - \nabla \boldsymbol{\lambda}:\boldsymbol{\sigma}(\mathbf{z}) - \nabla \mathbf{z}:\boldsymbol{\sigma}(\boldsymbol{\lambda}) \qquad (C.27)$$

$$\Theta = \frac{\partial^2 V_j}{\partial x_i \partial x_i} W_j \qquad (C.28)$$

Plugging Equations (C.17)-(C.28) to (C.16) results in the volume form of $<D<D\Psi;\mathbf{V}>;\mathbf{W}>$:

$$\langle D\langle D\psi;\mathbf{V}\rangle;\mathbf{W}\rangle = \iiint_{\Omega_\eta} \left[\left(\delta(\mathbf{x}-\mathbf{X})^T \mathbf{z}\right)\nabla \mathbf{V} + \mathbf{M}\right]:\nabla \mathbf{W} d\Omega_\eta + \iiint_{\Omega_\eta} \left[\delta(\mathbf{x}-\mathbf{X})^T \mathbf{z} + \mathrm{N}\right]\Theta d\Omega_\eta \qquad (C.29)$$

where the term **M** needs not to be calculated, since it is dropped when transforming to the boundary form by Guillaume-Masmoudi lemma, see Section 4.3. In the end, the second-order porosity sensitivity $<D^2\Psi:(\mathbf{V},\mathbf{W})>$ is available by combining Equation (C.16) with Equations (C.26) and (C.29).

**Appendix D. Exterior Formulation**

Similar to the derivation of exterior formulation for the primary solution (**z**) in Section 5.2, the exterior solutions of adjoints ($\boldsymbol{\lambda}_E$) is computed by solving the BVP in Equation (D.1) with the stress and strain fields defined in Equations (D.2) and (D.3).

$$\begin{cases} -\nabla \cdot \boldsymbol{\sigma}(\boldsymbol{\lambda}_\mathbf{E}) = \mathbf{0} & \mathbf{X} \in R^n - \Omega_p^1 \\ \boldsymbol{\lambda}_\mathbf{E} = \mathbf{0} & \mathbf{X} \to \infty \\ \boldsymbol{\sigma}(\boldsymbol{\lambda}_\mathbf{E}) \cdot \mathbf{n} = -\boldsymbol{\sigma}_\mathbf{0} \cdot \mathbf{n} & \mathbf{X} \in \Gamma_p^1 \end{cases} \quad (D.1)$$

$$\boldsymbol{\sigma}(\boldsymbol{\lambda}_\eta(\mathbf{x})) = \boldsymbol{\sigma}(\boldsymbol{\lambda}_0) + \boldsymbol{\sigma}(\boldsymbol{\lambda}_\mathbf{E}(\mathbf{X})) \quad (D.2)$$

$$\boldsymbol{\varepsilon}(\boldsymbol{\lambda}_\eta(\mathbf{x})) = \boldsymbol{\varepsilon}(\boldsymbol{\lambda}_0) + \boldsymbol{\varepsilon}(\boldsymbol{\lambda}_\mathbf{E}(\mathbf{X})) \quad (D.3)$$

where the index '1' indicates the full-size pore, **X** is the material point on the original porous domain with pore surface $\Gamma_p^1$, $\boldsymbol{\sigma}$ and $\boldsymbol{\varepsilon}$ are the stress and strain, $\boldsymbol{\lambda}_0$, $\boldsymbol{\lambda}_E$ and $\boldsymbol{\lambda}_\eta$ are the adjoints computed on the reference domain, from exterior formulation, and on the parameterized domain, respectively, and η is the pore shape parameter. The multiplier $\boldsymbol{Q}_\eta$, on the parameterized pore surface ($\mathbf{x} \in \Omega_p^\eta$) is computed by the BVP:

$$\begin{cases} -\nabla \cdot \boldsymbol{\sigma}(\mathbf{Q}_\eta(\mathbf{x})) = \mathbf{0} & \mathbf{x} \in \Omega - \Omega_p^\eta \\ \mathbf{Q}_\eta(\mathbf{x}) = \mathbf{0} & \mathbf{x} \in \Gamma^h \\ \boldsymbol{\sigma}(\mathbf{Q}_\eta(\mathbf{x})) \cdot \mathbf{n} = \mathbf{0} & \mathbf{x} \in \Gamma^s \\ \boldsymbol{\sigma}(\mathbf{Q}_\eta(\mathbf{x})) \cdot \mathbf{n} = -\nabla \cdot \mathbb{C} : \left( \frac{1}{2}(\mathbf{A} + \mathbf{A}^\mathbf{T}) \right) & \mathbf{x} \in \Gamma_p^\eta \end{cases} \quad (D.4)$$

where $\Gamma^s$ and $\Gamma^h$ are the boundaries prescribed with Neumann and Dirichlet boundary conditions, respectively, and the term **A** is:

$$\mathbf{A} = \mathbf{n} \otimes (\nabla \mathbf{z}_\eta^\mathbf{T}(\mathbf{x})\mathbf{V}) \quad (D.5)$$

For the assumed linear mapping in Equation (3.2), we have:

$$\boldsymbol{\sigma}(\mathbf{Q}_\eta(\mathbf{x})) = \boldsymbol{\sigma}(\mathbf{Q}(\mathbf{X}))/\eta \quad (D.6)$$

$$\nabla \mathbf{Q}_\eta(\mathbf{x}) = \nabla \mathbf{Q}(\mathbf{X})/\eta \quad (D.7)$$

By combining the Equations (D.6) and (D.7) with Equation (D.4), we have:

$$\begin{cases} -\nabla \cdot \boldsymbol{\sigma}(\mathbf{Q}(\mathbf{X})) = \mathbf{0} & \mathbf{X} \in \Omega - \Omega_p^1 \\ \mathbf{Q}(\mathbf{X}) = \mathbf{0} & \mathbf{X} \in \Gamma^h \\ \boldsymbol{\sigma}(\mathbf{Q}(\mathbf{X})) \cdot \mathbf{n} = \mathbf{0} & \mathbf{X} \in \Gamma^s \\ \boldsymbol{\sigma}(\mathbf{Q}(\mathbf{X})) \cdot \mathbf{n} = -\nabla \cdot \mathbb{C} : \left( \frac{1}{2}(\overline{\mathbf{A}} + \overline{\mathbf{A}}^{\mathrm{T}}) \right) & \mathbf{X} \in \Gamma_p^1 \end{cases} \tag{D.8}$$

with

$$\overline{\mathbf{A}} = \mathbf{n} \otimes (\nabla \mathbf{z}^{\mathrm{T}}(\mathbf{X})\mathbf{V}) \tag{D.9}$$

Let $\boldsymbol{Q}_E(X) = \boldsymbol{Q}(X)$, and approximate the solutions of Equation (D.8) by an exterior Neumann formulation as:

$$\begin{cases} -\nabla \cdot \boldsymbol{\sigma}(\mathbf{Q}_E(\mathbf{X})) = \mathbf{0} & \mathbf{X} \in R^n - \Omega_p^1 \\ \boldsymbol{\sigma}(\mathbf{Q}_E(\mathbf{X})) \cdot \mathbf{n} = -\nabla \cdot \mathbb{C} : \left( \frac{1}{2}(\overline{\mathbf{A}} + \overline{\mathbf{A}}^{\mathrm{T}}) \right) & \mathbf{X} \in \Gamma_p^1 \\ \mathbf{Q}_E(\mathbf{X}) = \mathbf{0} & \mathbf{X} \to \infty \end{cases} \tag{D.10}$$

Then, the stress and displacement gradients of $\boldsymbol{Q}_\eta$ can be approximated by:

$$\boldsymbol{\sigma}(\mathbf{Q}_\eta(\mathbf{x})) = \boldsymbol{\sigma}(\mathbf{Q}_E(\mathbf{X}))/\eta \tag{D.11}$$

$$\nabla \mathbf{Q}_\eta(\mathbf{x}) = \nabla \mathbf{Q}_E(\mathbf{X})/\eta \tag{D.12}$$

Similarly, the stress and displacement gradients of the multiplier $\boldsymbol{P}_\eta$ in Equation (C.15) are approximated explicitly as functions of the pore shape parameter $\eta$ as:

$$\boldsymbol{\sigma}(\mathbf{P}_\eta(\mathbf{x})) = \boldsymbol{\sigma}(\mathbf{P}_E(\mathbf{X}))/\eta \tag{D.13}$$

$$\nabla \mathbf{P}_\eta(\mathbf{x}) = \nabla \mathbf{P}_E(\mathbf{X})/\eta \tag{D.14}$$